\documentclass{aa}
\usepackage{babel}
\usepackage{graphicx}
\usepackage{newtxtext,newtxmath}
\usepackage{lipsum}
\usepackage{lscape}
\usepackage{placeins}
\usepackage{color}
\usepackage{textcomp}
\def\herschel{{\it Herschel}}
\def\spitzer{{\it Spitzer}}
\def\hubble{{\it HST}}
\def\kin{Kinemetry}
\def\cifull{[C\,{\sc i}] $^3P_2 \longrightarrow ^3P_1$}
\def\cilong{[C\,{\sc i}] J:2--1}
\def\cishort{[C\,{\sc i}] 2--1}
\def\co7-6{CO J:7--6}
\def\h2ofull{H$_2$O 2$_{11} - 2_{02}$}
\def\water{H$_2$O}

\def\350mic{350\,$\mu$m}
\def\micron{~$\mu$m}

\def\msun{\hbox{$\rm ~M_{\odot}$}}

\def\arcdeg{$^{\circ}$}
\def\H0{{\rm ~km\,s^{-1}\,Mpc^{-1}}}

\def\kms{km\,s$^{-1}$}
\def\src{H-ATLAS\,J084933.4+021443}

\begin{document}

    \title{Resolved Schmidt-Kennicutt relation in a binary hyperluminous infrared galaxy at $z=2.41$}
    \subtitle{ALMA observations of H-ATLAS\,J084933.4+021443}

    \author{
        Jonathan~S.~G\'omez            \inst{\ref{Madrid}, \ref{CIAFF}}   \and
        Hugo~Messias                   \inst{\ref{ALMA}, \ref{ESO-Chile}} \and
        Neil~M.~Nagar                  \inst{\ref{UdeC}}                  \and
        Gustavo~Orellana-Gonz\'alez    \inst{\ref{ANID}}                  \and
        R.\,J.~Ivison                  \inst{\ref{ESO-Germany}}           \and
        Paul~van~der~Werf              \inst{\ref{Leiden Obs}}
        }

    \institute{
        Departamento de F\'isica Te\'orica, Facultad de Ciencias, Universidad Aut\'onoma de Madrid, 28049 Cantoblanco, Madrid, Spain.
        \label{Madrid}\\
        \email{j.s.gomez.u@gmail.com}
        \and 
        Centro de Investigaci\'on Avanzada en F\'isica Fundamental (CIAFF), Facultad de Ciencias, Universidad Aut\'onoma de Madrid, 28049 Madrid, Spain. \label{CIAFF}
        \and
        Astronomy Department, Universidad de Concepci\'on, Concepci\'on, Chile. \label{UdeC}
        \and
        Joint ALMA Observatory, Alonso de C\'ordova 3107, Vitacura 763-0355, Santiago, Chile. \label{ALMA}
        \and
        European Southern Observatory, Alonso de C\'ordova 3107, Vitacura, Casilla 19001, Santiago de Chile, Chile. \label{ESO-Chile}
        \and
        Fundación Chilena de Astronomía, Santiago, Chile. \label{ANID}
        \and
        European Southern Observatory, Karl-Schwarzschild-Stra{\ss}e-2, 85748 Garching bei M{\"u}nchen, Germany. \label{ESO-Germany}
        \and
        Leiden Observatory, Leiden University, P.O. Box 9513, NL-2300 RA Leiden, The Netherlands. \label{Leiden Obs}
    }

    \abstract
        {}
        {Hyperluminous infrared galaxies (HyLIRGs; star-formation rates of up to $\approx 1000$ \msun\,yr$^{-1}$) -- while rare -- provide crucial long-lever-arm constraints on galaxy evolution. \src, a $z=2.41$ binary HyLIRG (galaxies `W' and `T') with at least two additional luminous companion galaxies (`C' and `M'), is thus an optimal test ground for studies of star formation and galaxy evolution during `cosmic noon'.}
        {We have used ALMA to obtain resolved imaging and kinematics of atomic and molecular emission lines, and rest-frame 340 to 1160~GHz continuum emission, for the galaxies W, T, M, and C.}
        {All four galaxies are spatially resolved in CO J:7--6, \cishort, \water, and the millimetre (mm) to sub-millimetre continuum, using circular apertures of $\sim$0\farcs3 (2.5\,kpc) in radius. Rotation-dominated gas kinematics are confirmed in W and T. The gas and continuum emission of galaxy T are extended along its kinematic minor axis, attributable to spatial lensing magnification. Spatially resolved sub-millimetre spectral energy distributions (SEDs) reveal that galaxy W is well fitted with greybody emission from dust at a single temperature over its full extent, despite hosting a powerful active galactic nucleus, while galaxy T requires an additional component of hotter nuclear dust and additional sources of emission in the millimetre. We confirm that \cilong\ can be used as a tracer of warm/dense molecular gas in extreme systems, though the \cilong/CO J:7--6 luminosity ratio increases sub-linearly. We obtain an exquisite resolved (2.5-kpc-scale) Schmidt-Kennicutt (SK) relationship for galaxies W and T, using both cold and warm/dense molecular gas. Gas exhaustion timescales for all apertures in W (T) are $\sim$50--100~Myr ($\sim$100--500~Myr). Both W and T follow a resolved SK relationship with a power-law index of $n\sim 1.7$, significantly steeper than the $n\sim 1$ found previously via cold molecular gas in nearby `normal' star-forming galaxies.}
        {}

    \keywords{Galaxies: formation -- galaxies: high-redshift -- infrared: galaxies -- infrared: jets and outflows -- radio continuum: galaxies -- submillimeter: galaxies}

    \authorrunning{Jonathan S. G\'omez et al.}
    \date{Accepted on September 23, 2025}
    \maketitle
    \nolinenumbers

    \section{Introduction}
        \label{sec : introduction}

        Sub-millimetre (sub-mm) surveys have greatly advanced our understanding of galaxy evolution by uncovering a population of heavily dust-obscured galaxies at high redshift, the so-called sub-mm galaxies (SMGs). Although the first such system to be discovered \citep{Ivison1998} was a hyperluminous infrared (IR) galaxy (HyLIRG; $L_{\rm IR}\, \geq \, 10^{13}$\,L$_\odot$, for which the IR luminosity is measured across $\lambda_{\mathrm{rest}} = 8$--$1000$\micron), the vast majority of the numerous SMGs uncovered thereafter were ultraluminous infrared galaxies (ULIRGs; $L_{\rm IR}\, \geq \, 10^{12}$\,L$_\odot$, forming stars at $\geq 100$\,M$_\odot$\,yr$^{-1}$; see, e.g. \citealt{Blain2002, Casey2014, Fudamoto2017}). The importance of SMGs in studies of galaxy formation and evolution has been highlighted by results from {\it Spitzer} and the {\it Herschel} Space Observatory, which show that SMGs contribute significantly to the total amount of star formation in the early Universe (e.g. \citealt{Magnelli2009}; \citealt{Glenn2010}).

        The IR luminosity of a HyLIRG implies an extreme star-formation rate (SFR; \citealt{Kennicutt1998}), SFR$\, \geq 10^3$ M$_{\odot}$ yr$^{-1}$, assuming that all of $L_{\rm IR}$ arises from star formation and adopting a `normal' \citep{Chabrier2003} stellar initial mass function (IMF), although \citet{Zhang2018} recently showed that the IMF in dusty starbursts is likely to be top heavy. Naively, this suggests starburst lifetimes of only $\sim 100$ Myr, unless star formation migrates around an extended gas reservoir. While rare and extreme, HyLIRGs are excellent laboratories for testing predictions from recent hydrodynamic simulations of galaxy evolution, including those of isolated and merging systems \citep[e.g.][]{Bahe2017, Barnes2017, Cote2018, Sukova2017, Schartmann2017, Lagos2018}. Their extreme physical conditions also allow us to investigate the validity of star-formation `laws' \citep[e.g.][]{Hayward2011}, and to study the role of feedback processes driven either by active galactic nucleus (AGN) activity \citep[e.g.][]{Bouche2010, Dave2011, Dave2012, Cicone2014} or by intense star formation \citep[e.g.][]{Shetty2012, Lilly2013} in regulating the evolution of galaxies.

        Of particular relevance in the study of dusty star-forming galaxies is the resolved Schmidt--Kennicutt (SK) relation \citep{Schmidt1959, Kennicutt1989}, i.e.\ the power-law relationship between the surface density of gas and star formation ($\mathrm{\Sigma_{SFR} \propto \Sigma_{H_2}^N}$) on sub-kiloparsec scales within a galaxy, and the use of sub-mm C\,{\sc i} lines as an alternative to CO \citep{FlowerAndLaunay1985, DownesAndSolomon1998, Yang2010, Bolatto2013, CarilliAndWalter2013, Rodriguez2014} or HCN \citep{Gao1997, GaoAndSolomon2004, Shimajiri2017, Oteo2017} to estimate the total molecular gas content of galaxies \citep{Walter2011, Israel2015, Jiao2017}. While the galaxy-integrated SK relation, i.e.\ the surface densities of the galaxy-integrated star formation and cold molecular gas mass (hereafter referred to as the galaxy-integrated SK relation), has been extensively studied --  for example by\ \citet{Young1986}, \citet{SolomonAndSage1988}, \citet{Buat1989}, \citet{GaoAndSolomon2004}, \citet{Bouche2007}, \citet{KrumholzAndThompson2007}, \citet{Daddi2010}, \citet{Genzel2010}, \citet{KennicuttAndEvans2012}, and \citet{Sharon2016} -- the resolved SK relation (at $\lesssim$1-kpc resolution within galaxies; hereafter the kpc-scale SK relation) has been constrained in relatively few nearby \citep{WongAndBlitz2002, Kennicutt2007, Bigiel2008, Leroy2008, Krumholz2009, Bigiel2010, Bigiel2011, Boquien2011, Momose2013, Roychowdhury2015} and high-redshift \citep{Freundlich2013, Thomson2015} galaxies. \citet{Bigiel2008} and \citet{Bigiel2011} found that the 1~kpc-scale resolved SK relation in their `normal' star-forming galaxies is consistent with an exponent of $N \sim 1$. In contrast, \citet{Momose2013} reported a super-linear slope ($N=1.3$, and even up to 1.8) for the resolved SK relation in their sample of nearby spirals. A super-linear slope ($N=1.5$) was also identified by \citet{Roychowdhury2015} in H\,{\sc i}-dominated regions of nearby spiral and dwarf galaxies.

        The \cifull\ and [C\,{\sc i}] $^3P_1 \longrightarrow ^3P_0$ lines -- both are required to determine the excitation temperature\footnote{Noting that \citet{Papadopoulos2022} found that average excitation conditions are often strongly sub-thermal, leading to LTE-assumed $T_{\rm gas}$ values clustering around $\approx 25$\,K, as reported in, for example, the work of \citet{Valentino2020}.} of C\,{\sc i} \citep{Stutzki1997} -- can be used to robustly determine the mass of the neutral carbon \citep{Stutzki1997, WeiB2003, WeiB2005}. In ultraviolet (UV) or cosmic ray-dominated regions, the (typically optically thin) C\,{\sc i} emission lines are expected to be primarily produced in the dissociated surfaces of molecular clouds, though observations show that they are present throughout the cloud \citep[e.g.][]{Glover2015} and it has been argued that they are a better tracer of total molecular gas mass than the CO line \citep{WeiB2003, WeiB2005, Bisbas2015, Glover2016}. The detection of CO and \cishort\ emission lines in multiple high-redshift galaxies, particularly those with spatially resolved data, enables comparative studies of these species as molecular gas tracers. Such comparisons are especially valuable in extreme environments in which different tracers may probe distinct phases of the interstellar medium. Molecular lines from species such as H$_2$O, HCN, and CS have a much higher critical density, and therefore probe the dense molecular star-forming phase. \citet{Omont2013} reported a relation between the far-infrared (FIR) and H$_2$O luminosities for a sample of high-redshift starburst galaxies. The H$_2$O detections for this sample are all associated with underlying FIR emission, implying that the H$_2$O emission traces star-forming regions. However, the H$_2$O molecules can also be excited in the dissipation of supersonic turbulence in molecular gas or by slow shocks (e.g. \citealt{FlowerAndPineau2010}). In the case of purely shock-excited H$_2$O, it is unlikely that underlying FIR emission would be detected in regions of strong H$_2$O emission (e.g. \citealt{Goicoechea2015, Anderl2013}).

        The HyLIRG HATLAS\,J084933.4+021443 (hereafter HATLAS\,J084933) at $z=2.41$, first identified in 2012, is now one of the few well-studied HyLIRGs \citep{Ivison2013} (hereafter I13). It has a relatively brief literature history: identified in the \herschel\ ATLAS imaging survey \citep{Eales2010} as a \350mic\ peaker, with subsequent CO J:1--0 spectroscopy using the Green Bank Telescope (GBT) constraining its redshift to 2.410 \citep{Harris2012}. I13 presented a comprehensive analysis of the molecular gas and rest-frame sub-mm emission of this source using the Jansky Very Large Array (CO J:1--0), CARMA (CO J:3--2), IRAM PdBI (CO J:4--3), together with continuum imaging from the SMA, \herschel, \spitzer, VISTA, and the \textit{Hubble} Space Telescope (\hubble), and optical spectroscopy from Keck. They determined that HATLAS\,J084933 comprises at least four starburst galaxies scattered across a $\sim 100$-kpc region at $z=2.41$. The two brightest galaxies, dubbed W and T, both HyLIRGs in their own right, are separated by $\sim 85$~kpc on the sky. Of these, T is amplified modestly by a foreground galaxy with a lensing magnification (in flux) of $\sim 2\times$. The molecular gas reservoirs of W and T, each $\sim 3$~kpc in size, are rotation-dominated and counter-rotate. These two galaxies have CO line strengths and widths typical of the brightest SMGs and lie among SMGs in the `global' (i.e.\ galaxy-wide) equivalent of the SK relation. The two other known galaxies in this system, dubbed M and C, are ULIRGs, though they lie relatively close to the HyLIRG cut-off. Their estimated molecular gas masses (but not their dynamical masses) are almost an order of magnitude lower than those of W and T.

        As a rare system hosting multiple IR-luminous starbursts at the same redshift, HATLAS\,J084933 provides a unique opportunity to probe the diversity of star-formation modes, interstellar medium (ISM) conditions, and gas kinematics within a single high-redshift environment. Its combination of extreme SFRs, resolved kinematics, and favourable lensing makes it an ideal laboratory for testing star-formation laws and gas tracers under the most intense conditions.

        With the aim of more comprehensively studying the physics in HATLAS\,J084933 -- particularly to probe the resolved SK relation at the highest SFRs and gas densities, but also to explore the resolved kinematics (rotation vs\ dispersion) and to search for evidence of outflows via kinematics and P~Cygni profiles (e.g. molecular outflows with velocities $\gtrsim$\,500\,km\,s$^{-1}$ as seen by \citealt{Cicone2014}, \citealt{Feruglio2010}) -- we have obtained new observations of this source using the Atacama Large Millimetre/sub-millimetre Array (ALMA). These new observations include resolved imaging of the CO J:3--2 and CO J:7--6 emission lines (tracers of the dense molecular gas), the \cifull\ fine-structure line ($\nu_{\mathrm{rest}} = 809.34$~GHz; hereafter \cishort), and the \h2ofull\ emission line ($\nu_{\mathrm{rest}} = 752.03$~GHz; hereafter H$_2$O). These lines are detected and resolved in all four principal galaxies of HATLAS\,J084933 -- W, T, C, and M -- significantly improving our understanding of the source compared to I13. We have also obtained resolved continuum imaging at rest-frame 341~GHz, 750~GHz, 808~GHz, and 1160~GHz, which we use to trace the resolved star-formation rate, dust mass, and temperature.

        In this work we present the first set of results based on the new ALMA data. In Sect.~2 we describe our ALMA sub-mm observations. The results of these observations are presented in Sect.~3. In Sect.~4 we analyse and discuss the results, and present our conclusions. We adopt a cosmology with $H_0 = 71$ km s$^{-1}$ Mpc$^{-1}$, $\Omega_m = 0.27$, and $\Omega_\Lambda = 0.73$, so that 1\arcsec\ corresponds to 8.25 kpc at $z = 2.41$.
        
    \section{Observations and data processing}

        HATLAS\,J084933 was observed by ALMA as part of Project 2012.1.00334.S (P.I.: G.~Orellana) during ALMA Cycle~2. We used the Common Astronomy Software Applications (CASA, version~4.4.0) for all data calibration and imaging steps, producing the final data cubes, continuum maps, emission-line-only cubes, and moment~0, 1, and~2 maps of the individual emission lines. Further processing and analysis were performed with our Interactive Data Language (IDL) and Python codes.
        
        \subsection{CO J:3--2 and rest-frame 341~GHz (880\micron) continuum imaging}

            Approximately 35\,min of on-source integration was obtained in Band~3 with 34 12-metre antennas on 30~August 2015. The Band~6 receivers were tuned to observe the CO J:3--2 line ($\nu_{\mathrm{rest}}=345.795991$\,GHz; \citealt{MortonAndNoreau1994}), redshifted to $\nu_{\mathrm{obs}}=101.465$\,GHz, in one of the four spectral windows (SPWs). The second SPW in the upper sideband (USB) was set to partially cover the CS J:7--6 line. The two SPWs in the lower sideband (LSB), which do not include any known strong lines, were used to detect continuum emission. To maximise sensitivity, the ALMA correlator was used in `continuum' or Time Division Multiplexing (TDM) mode, which provided a spacing of $\sim 46$\,km\,s$^{-1}$ per channel and a total velocity coverage of $\sim 5000$\,\kms\ per SPW.
            
            Using a Briggs weighting with a robust parameter of +0.5, and the intrinsic spectral resolution, the CO line was imaged with a synthesised (FWHM) beam of $1\farcs13 \times 0\farcs51$ with a position angle (P.A.) of 115\arcdeg. The r.m.s.\ noise was $\sigma = 0.27$\,mJy\,beam$^{-1}$ in each 46\,km\,s$^{-1}$ channel. Line-free channels were used to create a continuum map at the observed-frame frequency of $\sim 100$\,GHz (rest-frame $\sim 341$\,GHz); here the synthesised beam was $1\farcs23 \times 0\farcs56$ FWHM (P.A.\ 116\arcdeg), and the r.m.s.\ noise was $\sigma = 0.08$\,mJy\,beam$^{-1}$.
        
        \subsection{CO J:7--6, \cifull, and rest-frame 808~GHz (370\micron) continuum imaging}
            \label{sec : co76, c1 and their continnum imaging}
            
            Approximately 21\,min of on-source integration time was obtained in Band~6 with 34 12-metre antennas in September 2015. The last scan of the observation was not on the phase calibrator, but rather on the target. Nevertheless, the target was bright enough to enable self-calibration, making that scan usable. Two overlapping SPWs in the USB were tuned to contiguously cover both the CO J:7--6 line ($\nu_{\mathrm{rest}} = 806.651801$\,GHz; \citealt{MortonAndNoreau1994}), redshifted to an observed frequency of $\sim 236$\,GHz, and the \cifull\ (\cishort) line ($\nu_{\mathrm{rest}} = 809.34197$\,GHz; \citealt{Muller2001}), redshifted to an observed frequency of $\sim 237$\,GHz. The other two SPWs were set to cover a \water\ line and neighbouring continuum (see Sect.~\ref{sec : h2o and it continnum imaging}). Once more, for maximum sensitivity, the correlator was set to TDM mode, resulting in a spectral spacing of 19.7\,km\,s$^{-1}$ per channel and a total velocity coverage of $\sim 2500$\,\kms\ per SPW for all SPWs.
            
            Image cubes of the CO J:7--6 and \cishort\ lines were made at the intrinsic spectral resolution (19.7\,km\,s$^{-1}$ channels) using natural weighting, resulting in a synthesised (FWHM) beam of $0\farcs27 \times 0\farcs24$ at P.A.\ 3.9\arcdeg. The r.m.s.\ noise in each channel was 0.31\,mJy\,beam$^{-1}$ for the CO J:7--6 and \cishort\ line cubes. Line-free channels were used to make a map of the continuum emission near an observed frequency of 237\,GHz (corresponding to a rest frequency of 808\,GHz). Here, maps made with a Briggs robust parameter of +0.5 resulted in a synthesised beam of $0\farcs26 \times 0\farcs24$ at P.A.\ 2.9\arcdeg and an r.m.s.\ noise of $\sigma = 0.24$\,mJy\,beam$^{-1}$.
        
        \subsection{\h2ofull\ imaging, and 750~GHz (400\micron) continuum imaging}
            \label{sec : h2o and it continnum imaging}
            
            In the science goal described in Sect.~\ref{sec : co76, c1 and their continnum imaging}, the two LSB SPWs were set to 220.538\,GHz, the redshifted frequency of the \h2ofull\ (H$_2$O) line ($\nu_{\mathrm{rest}} = 752.033143$\,GHz; \citealt{Dionne1980}), and the neighbouring continuum. The spectral spacing was 21.2\,km\,s$^{-1}$ per channel. The image cube of the \h2ofull\ line, made with natural weighting, had an r.m.s.\ noise of 0.34\,mJy\,beam$^{-1}$ per native channel and a synthesised (FWHM) beam size of $0\farcs29 \times 0\farcs26$ at P.A.\ 13.7\arcdeg. Line-free channels in these two LSB SPWs were used to make a map of the continuum emission near an observed frequency of 220\,GHz (corresponding to a rest-frame frequency of 750\,GHz). Here, maps made with a Briggs robust parameter of +0.5 resulted in a synthesised beam of $0\farcs28 \times 0\farcs26$ at P.A.\ 179\arcdeg and an r.m.s.\ noise of $\sigma = 0.17$\,\textmu Jy\,beam$^{-1}$.

        \subsection{Observed frame 340~GHz (rest-frame 1160 GHz or 260\micron) continuum imaging}

            Approximately 53\,min of on-source integration was obtained with 38 working 12-metre antennas on 5~June 2015. The spectral set-up was designed to detect four different \water\ lines, one in each SPW, none of which were detected. This dataset was thus used only to obtain a continuum image near an observed-frame frequency of 340\,GHz (corresponding to a rest-frame frequency of 1160\,GHz). The receivers were tuned such that the USB was centred at 1160\,GHz (260\micron). A weighting scheme with a Briggs robust parameter of +0.5 resulted in a synthesised beam of $0\farcs34 \times 0\farcs3$ (P.A.\ 83.6\arcdeg) and a noise level of $\sigma = 0.26$\,mJy\,beam$^{-1}$ in the continuum image.
            
    \section{Analysis methods}
        \label{sec : analysis methods}

        In this section we describe the methods used to analyse the multi-line ALMA data of the four galaxies in the HATLAS\,J084933 system. Our approach is structured to extract both global and spatially resolved properties from the observed continuum and emission-line data, enabling an in-depth investigation of the gas dynamics and star-formation activity in each galaxy.

        We first describe the construction of a toy rotation model, which serves as the basis for interpreting the observed velocity fields and identifying possible outflow signatures. We then detail how this model is used to separate blended emission lines, extract galaxy-integrated fluxes, and derive physical quantities such as dynamical and molecular-gas masses. Finally, we present the continuum and line maps that underpin our spatially resolved analysis.

        \subsection{Rotational model construction}
            \label{sec: Rotational Model Construction}
        
            \begin{figure*}[ht]
                \hspace{-0.8cm}  \includegraphics[scale=0.44]{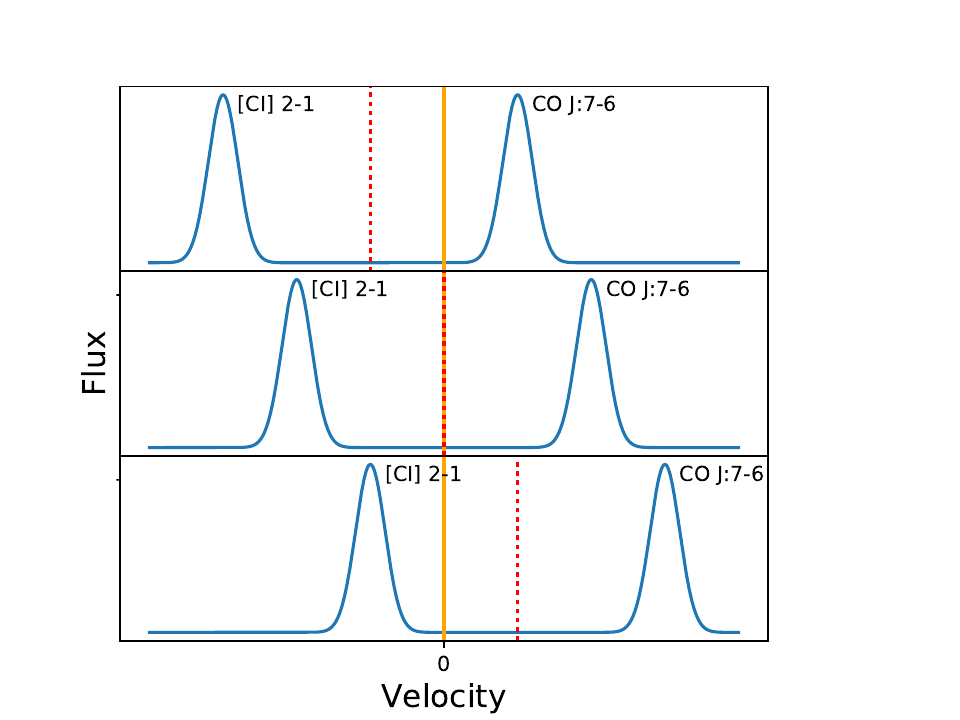}
                \hspace{-1.5cm}  \includegraphics[scale=0.44]{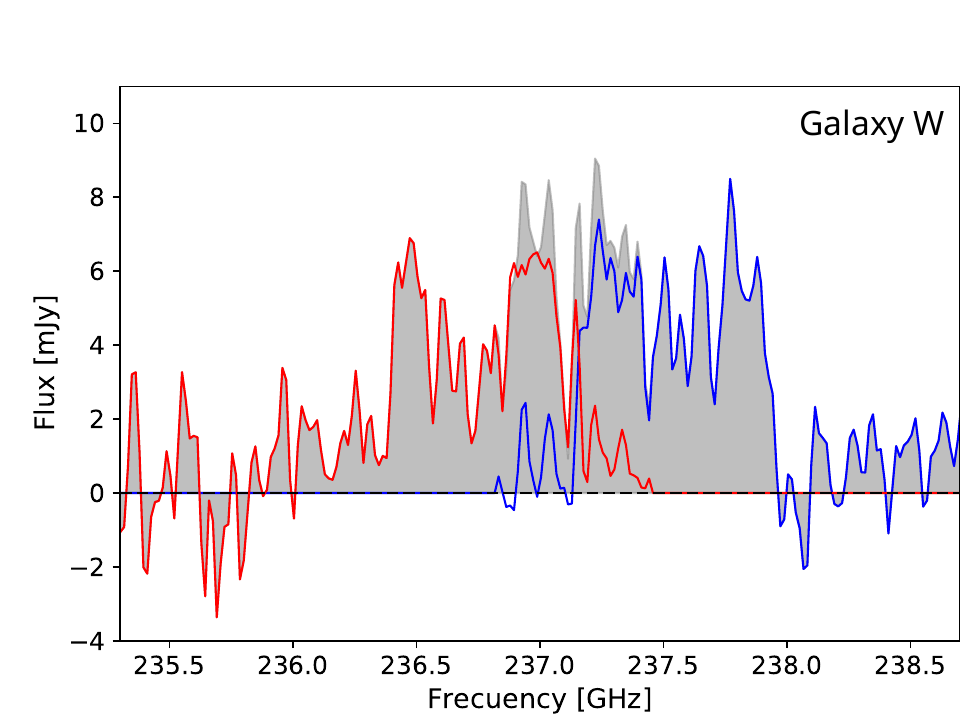}
                \hspace{-0.16cm} \includegraphics[scale=0.44]{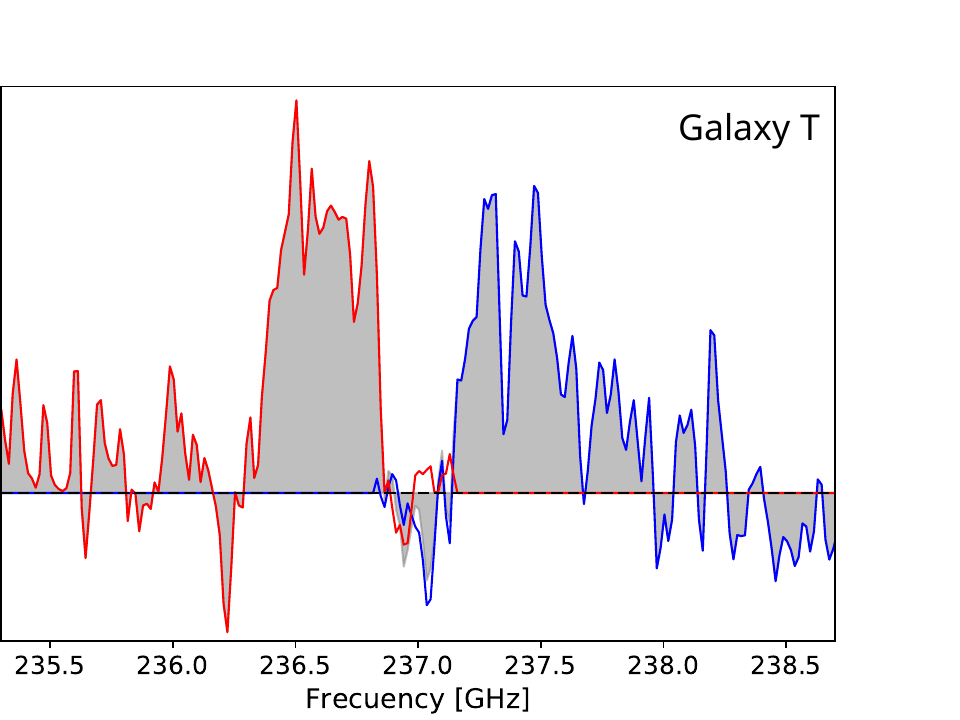}
                \caption{Left panel: Illustration of the method used to separate the \co7-6\ and \cishort\ lines in individual channels of the datacube (see text). The dashed red line shows our choice for the dividing frequency (based on the fitted gas rotation model) and the solid yellow line shows the systemic velocity of the galaxy (see Sect.~\ref{sec: Galaxy-integrated spectral properties}). Middle and right panels: Observed (filled grey regions) and rotation-model-separated profiles (see text) of the CO J:7--6 (red) and \cishort\ (blue) emission lines in galaxy W (middle panel) and galaxy T (right panel).}
                \label{fig : rotation-model-separated profiles}
            \end{figure*}

            Although velocity fields in high-redshift galaxies could, in principle, be interpreted as arising from either rotation or outflows within the disc, in this sub-section we focus exclusively on modelling ordered rotation. This rotational model serves as the basis for identifying and interpreting velocity-field patterns and potential outflows in the following sub-sections.

            The velocity fields of all strongly detected emission lines in galaxies W and T were input into the \kin\ package \citep{Krajnovic2006} to constrain the kinematic centre, position angle (PA) of the line of nodes, inclination, and rotation curve of each galaxy, assuming that the velocity field is rotation dominated. The initial guess for the nuclear position was set to the peak of the rest-frame 1160\,GHz continuum (see caption of Fig.~\ref{fig : 1160ghz fluxes}). The resulting kinematic parameters derived with \kin\ are listed in Tables~\ref{tab : w-kinemetry} and \ref{tab : t-kinemetry}. In these tables, $V_{\mathrm{asym}}$ corresponds to the inclination-corrected asymptotic rotational velocity.

            \begin{table}[ht]
                \caption{Results of \kin\ modelling of galaxy W.}
                \centering
                \begin{tabular}{c c c c}
                \hline \hline
                Property                          & CO J:3--2    & CO J:7--6    & \cifull      \\
                \hline
                PA.                               & $45^{\circ}$ & $55^{\circ}$ & $55^{\circ}$ \\
                Inclination                       & $48^{\circ}$ & $48^{\circ}$ & $48^{\circ}$ \\
                $V_{\mathrm{asym}}$ (km s$^{-1}$) & 464          & 525          & 525          \\
                \hline
                \end{tabular}
                \newline
                {
                \footnotesize{PA.: Position angle (N to E). Inclination: Angle between the disc and the plane of the sky, where 0$^{\circ}$ corresponds to face-on and 90$^{\circ}$ to edge-on. V$_{\rm asym}$: Peak to peak de-projected rotation velocity.}
                }
                \label{tab : w-kinemetry}
            \end{table}
    
        \begin{table}[ht]
                \caption{Results of \kin\ modelling of galaxy T.}
                \centering
                \begin{tabular}{c c c c}
                \hline \hline
                Property                          & CO J:3--2     & CO J:7--6     & \cifull       \\
                \hline
                PA.                               & $144^{\circ}$ & $136^{\circ}$ & $135^{\circ}$ \\
                Inclination                       & $49^{\circ}$  & $54^{\circ}$  & $49^{\circ}$  \\
                $V_{\mathrm{asym}}$ (km s$^{-1}$) & 298           & 241           & 298           \\
                \hline
                \end{tabular}
                \newline
                {
                \noindent Rows are as in Table~\ref{tab : w-kinemetry}.
                }
                \label{tab : t-kinemetry}
        \end{table}
            
            For each galaxy, the \kin-derived PA and inclination are consistent for the CO J:7--6 and \cishort\ lines. In contrast, the CO J:3--2 results show larger deviations, likely due to the significantly lower spatial resolution of the observations at this line.

            To facilitate comparisons with the observed velocity fields and aid in identifying deviations from pure rotation, we constructed a simplified toy model of ordered rotation. The model consists of solid-body rotation at small radii, transitioning to a flat rotation curve at larger radii:
            \begin{equation}
            V_{\mathrm{radial}} = \left\{
                \begin{array}{ll}
                V_{\mathrm{slope}} \cdot r \cdot \cos(\phi) \cdot \sin(i)                 & \mathrm{if\ } r \le r_{\mathrm{flat}}, \\
                V_{\mathrm{slope}} \cdot r_{\mathrm{flat}} \cdot \cos(\phi) \cdot \sin(i) & \mathrm{if\ } r >  r_{\mathrm{flat}} , \\
                \end{array}
            \right.
            \end{equation}
            where $V_{\mathrm{radial}}$ is the observed radial velocity in km\,s$^{-1}$, $V_{\mathrm{slope}}$ is the slope of the solid-body rotation in the inner region (km\,s$^{-1}$\,kpc$^{-1}$), $r$ (kpc) and $\phi$ (degrees) are the polar coordinates in the galaxy disc, and $r_{\mathrm{flat}}$ is the radius beyond which the rotation curve flattens.

            The model was initially constrained using the \kin-derived parameters in Tables~\ref{tab : w-kinemetry} and \ref{tab : t-kinemetry}, adopting the kinematic centre determined by \kin\ for each galaxy. We then varied $V_{\mathrm{slope}}$ and $r_{\mathrm{flat}}$ to obtain the best agreement — via visual inspection — between the model predictions and both the observed velocity profile along the major axis and the \kin-derived rotation curves.

            This toy model was used in all subsequent analyses (e.g. residual velocity maps), as it provides a smooth representation of the velocity field that closely follows the \kin-derived kinematics while suppressing small-scale variations that are likely caused by resolution or signal-to-noise limitations rather than by true physical substructures.

        \subsection{Separation of CO J:7--6 and \cishort\ lines using the rotation model}
            \label{sec: line separation method}

            \begin{table*}[ht]
                \caption{HATLAS J084933.4+021443: Line fluxes and luminosities, observed continuum fluxes, and derived properties.}
                \centering
                \begin{tabular}{l l c c c c}
                \hline\hline
                Property                                               & unit of measurement           &        W        &        T       &        C       &       M          \\
                \hline\hline
                $S_{\mathrm{CO\, J:1-0}}^a$                            & [Jy km s$^{-1}$]              & 0.49$\pm$0.06   & 0.56$\pm$0.07  & 0.057$\pm$0.013 & 0.079$\pm$0.014 \\
                $S_{\mathrm{CO\, J:3-2}}$                              & [Jy km s$^{-1}$]              & 4.07$\pm$0.21   & 4.18$\pm$0.16  & 0.98$\pm$0.09   & 0.62$\pm$0.08   \\
                $S_{\mathrm{CO\, J:7-6}}$                              & [Jy km s$^{-1}$]              & 2.57$\pm$0.24   & 2.71$\pm$0.17  & 0.53$\pm$0.06   & 0.55$\pm$0.05   \\
                $S_{\mathrm{[C\,{i}]\, 2-1}}$                          & [Jy km s$^{-1}$]              & 2.34$\pm$0.25   & 1.16$\pm$0.13  & 0.24$\pm$0.04   & -               \\
                $S_{\mathrm{H_2O}}$                                    & [Jy km s$^{-1}$]              & 0.80$\pm$0.11   & 1.01$\pm$0.10  & -               & -               \\
                \hline
                $L'^a_\mathrm{CO\, J:1-0}$                             & [$10^9$ K km s$^{-1}$ pc$^2$] & 138$\pm$17      & 157$\pm$20     & 16.0$\pm$3.6    & 22.2$\pm$3.9    \\
                $L'_\mathrm{CO\, J:3-2}$                               & [$10^9$ K km s$^{-1}$ pc$^2$] & 126.9$\pm$6.5   & 130.1$\pm$5.0  & 30.5$\pm$2.9    & 19.3$\pm$2.5    \\
                $L'_\mathrm{CO\, J:7-6}$                               & [$10^9$ K km s$^{-1}$ pc$^2$] & 14.8$ \pm$1.4   & 15.6$ \pm$1.0  & 3.05$\pm$0.37   & 3.15$\pm$0.28   \\
                $L'_\mathrm{[C\,{i}] 2-1}$                             & [$10^9$ K km s$^{-1}$ pc$^2$] & 13.4$ \pm$1.4   & 6.66$ \pm$0.76  & 1.36$\pm$0.22   & -              \\
                $L'_\mathrm{H_2O}$                                     & [$10^9$ K km s$^{-1}$ pc$^2$] & 5.29$ \pm$0.7   & 6.66$\pm$0.64  & -               & -               \\
                \hline
                $\mathrm{S_{\nu obs 100GHz}}$                          & [mJy]                         & 0.16 $\pm$0.02  & 0.15$\pm$0.02  & -               & -               \\
                $\mathrm{S_{\nu obs 220GHz}}$                          & [mJy]                         & 5.18 $\pm$0.14  & 3.67$\pm$0.13  & 0.41$\pm$0.04   & 0.36$\pm$0.04   \\
                $\mathrm{S_{\nu obs 237GHz}}$                          & [mJy]                         & 5.75 $\pm$0.19  & 4.33$\pm$0.20  & 0.39$\pm$0.06   & 0.51$\pm$0.07   \\
                $\mathrm{S_{\nu obs 340GHz}}$                          & [mJy]                         & 21.9 $\pm$0.19  & 21.9$\pm$0.25  & 2.81$\pm$0.14   & 1.89$\pm$0.21   \\
                \hline
                log $M_\mathrm{dyn}$ (from CO J:3-2)                   & [$M_\odot$]                   & 11.86$\pm$0.7   & 11.51$\pm$0.9  & 10.55$\pm$0.13  & 10.22$\pm$0.16  \\
                log $M_\mathrm{dyn}$ (from CO J:7-6)                   & [$M_\odot$]                   & 10.32$\pm$0.6   & 9.98 $\pm$0.6  & 8.33$\pm$0.08   & 8.51$\pm$0.09   \\
                (T$_{\rm ex}$=30~K): log$M_\mathrm{CI}$                & [$M_\odot$]                   & 6.87$\pm$0.08   & 6.57$\pm$0.06  & 5.88$\pm$0.07   & -               \\
                (T$_{\rm ex}$=40~K): log$M_\mathrm{CI}$                & [$M_\odot$]                   & 6.74$\pm$0.07   & 6.44$\pm$0.05  & 5.75$\pm$0.06   & -               \\
                log $M_\mathrm{\rm H_2}$  (from CI; T$_{\rm ex}$=40~K) & [$M_\odot$]                   & 11.44$\pm$0.14  & 11.14$\pm$0.11 & 10.45$\pm$0.13  & -               \\
                
                log $L_{{IR}}^a$                                       & [$L_\odot$]                   & 13.52$\pm$0.04  & 13.16$\pm$0.05 & 12.9$\pm$0.2    & 12.8$\pm$0.2    \\
                $\mathrm{T_{dust}^a}$                                  & [K]                           & 39.8$\pm$1.0    & 36.2$\pm$1.1   & -               & -               \\
                log $M_\mathrm{H_2}$ (from CO J:1-0)$^a$               & [$M_\odot$]                   & 11.04$\pm$0.12  & 10.92$\pm$0.13 & 10.25$\pm$0.15  & 10.11$\pm$0.14  \\   
                \hline
                \end{tabular}
                
                 $^a$ : from Ivison et al. (2013). \\
                \label{tab : global}
            \end{table*}

            The emission lines of CO J:3--2, CO J:7--6, \cishort, and \h2ofull\ are detected in the four previously identified galaxies of the HATLAS\,J084933 system: W, T, M, and C. Of these, galaxy C shows the weakest line emission.

            The observed galaxy-integrated spectra of CO J:7--6 and \cishort\ in galaxies W and T are shown in Fig.~\ref{fig : rotation-model-separated profiles}. Unfortunately, the relatively close rest-frame frequencies of these two lines ($\Delta \nu = 2.7$\,GHz, corresponding to $\Delta v \approx 1050$\,km\,s$^{-1}$), combined with the broad intrinsic line profiles in galaxy W, result in a significant overlap in their galaxy-integrated spectra. A similar, although less severe, situation is observed in galaxy T.

            Given that CO J:7--6 and \cishort\ display similar ordered velocity fields, and assuming that these fields are dominated by rotation (see Sect.~\ref{sec: Rotational Model Construction}), we can exploit this property to separate the emission into their respective spectral components. Specifically, the two lines can be cleanly disentangled in individual velocity channels of the data cube (channel width 19.7\,km\,s$^{-1}$), since the typical line widths in any given channel are $\sim$100--250\,km\,s$^{-1}$, while the velocity separation between the two lines is $\sim$1000\,km\,s$^{-1}$. This clean separation was visually verified in the data cube.

            We therefore used the rotation-only velocity models described in Sect.~\ref{sec: Rotational Model Construction} to predict, for each spatial pixel, the velocity (or spectral channel) that most effectively separates the CO J:7--6 and \cishort\ emission. The left panel of Fig.~\ref{fig : rotation-model-separated profiles} illustrates the resulting dividing frequency (dashed red line), as predicted by the best-fit rotation model. The rotation-model-separated profiles are shown in blue and red in the middle and right panels of Fig.~\ref{fig : rotation-model-separated profiles} for galaxies W and T, respectively.

            Note that this method assumes that all emitting gas is participating in ordered rotation. A weak outflow component with projected velocities of $\gtrsim$500\,km\,s$^{-1}$ could contaminate the separation and may not be reliably disentangled by this approach.

            Finally, the galaxy-integrated profiles of all detected emission lines in galaxies W and T — along with those for galaxies M and C (where de-blending was not required) — are presented in Sect.~\ref{sec: Continuum and Emission-Line Maps}.

        \subsection{Galaxy-integrated measurements and derived quantities}
            \label{sec: Galaxy-integrated Measurements and Derived Quantities}

            \begin{figure*}[ht]
                \centering
                \includegraphics[scale=0.37]{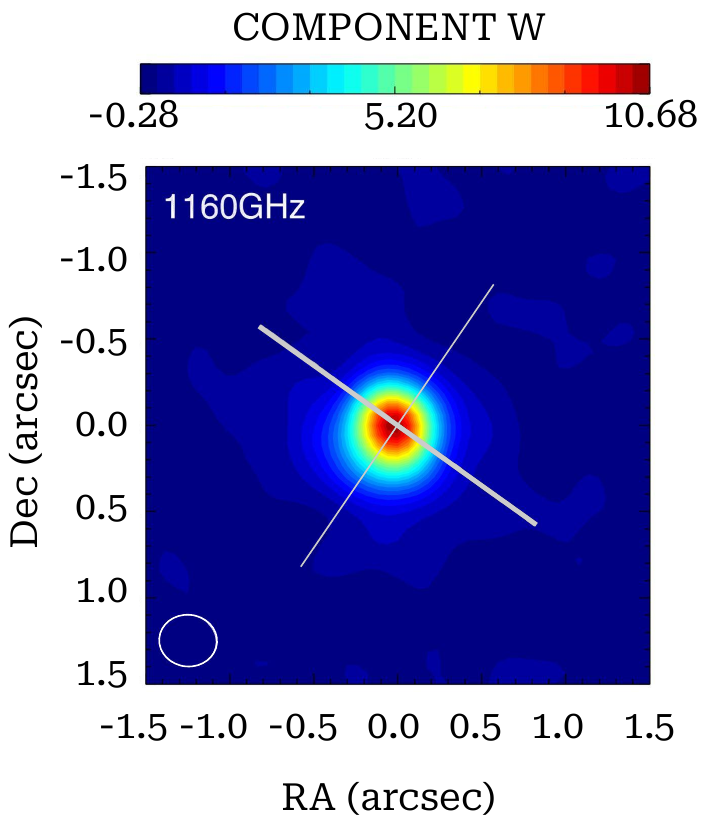}
                \includegraphics[scale=0.37]{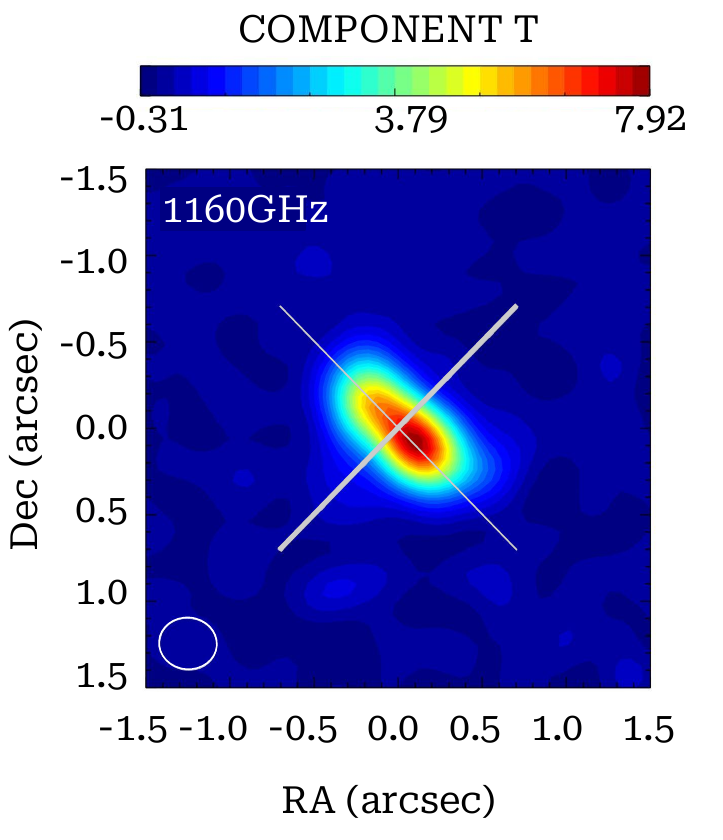}
                \includegraphics[scale=0.37]{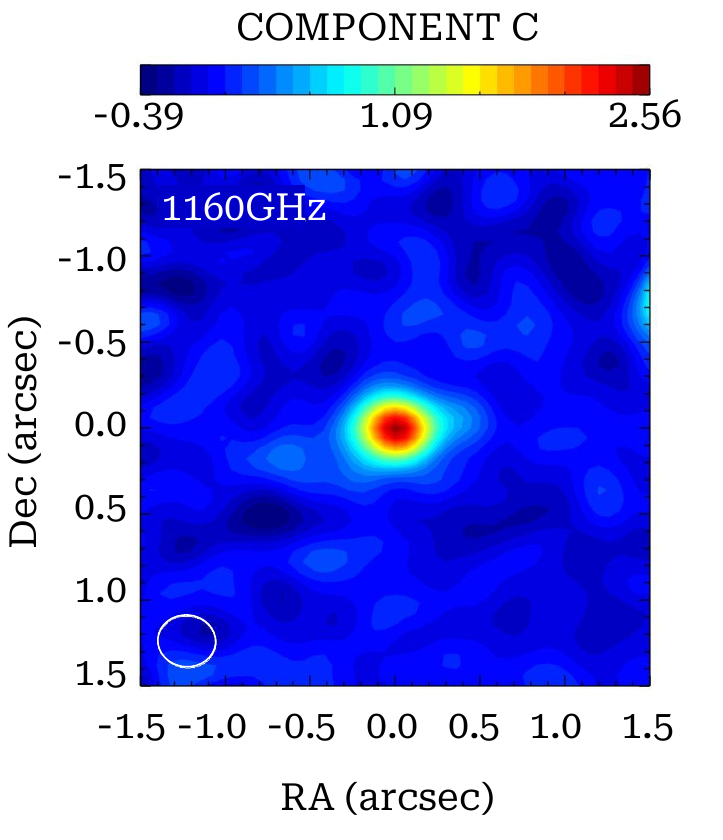}
                \includegraphics[scale=0.37]{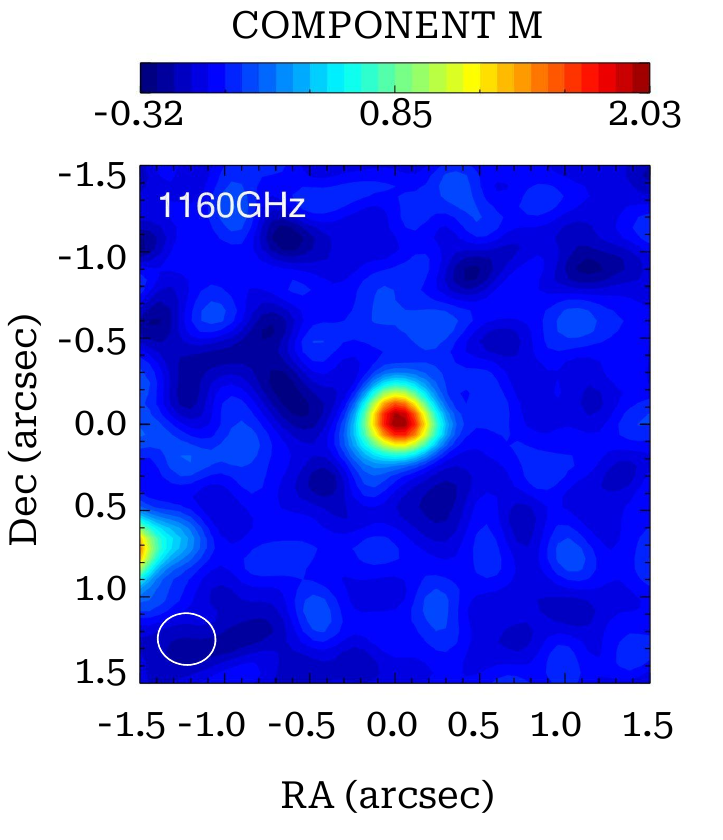}
                \caption{Maps of the observed-frame 340-GHz (corresponding to rest-frame 260\micron) continuum emission in (left to right) galaxies W, T, C and M. Continuum fluxes are in mJy $\mathrm{beam^{-1}}$ following the colour bar above each panel. Each panel is 3\arcsec\ $\times$ 3\arcsec\ in size and the axes, in arcsec, are centred on the kinematic centre of each galaxy, as obtained from \textit{Kinemetry}. These kinematic nuclear positions, hereafter used as the galaxy positions, are: 
                W: 08:49:33.685, +02:14:44.680; 
                T: 08:49:32.947, +02:14:39.696; 
                C: 08:49:33.908, +02:14:44.860; 
                M: 08:49:33.795, +02:14:45.595. 
                For galaxies W and T, the kinematic major and minor axes are shown with  thick and thin grey lines, respectively.
                }
                \label{fig : 1160ghz fluxes}
            \end{figure*}

            Galaxy-integrated measurements of the continuum flux densities, emission-line fluxes and luminosities, dynamical-mass estimates, and molecular-gas masses derived from our new dataset are listed in Table~\ref{tab : global} for each galaxy: W, T, C, and M. For easy reference, the estimated dust temperatures and total IR luminosities derived from the global SED modelling of I13 are also included. We rely on the I13 modelling as it remains the most comprehensive multi-wavelength analysis available for these sources, incorporating data from \textit{Herschel}, \textit{Spitzer}, SMA, and other key facilities. No substantial new photometric constraints have become available since.

            The quantities listed in Table~\ref{tab : global} were derived as follows. The dynamical mass was calculated under the assumption that the galaxy velocity field is rotation dominated (see Sect.~\ref{sec: Rotational Model Construction}):
            \begin{equation}
                M_\mathrm{dyn} = \frac{R \times (v_\mathrm{max}/ \sin i)^2}{G} = \frac{R \times V_{\mathrm{asym}}^2}{G},
            \end{equation}
            where $G$ is the gravitational constant, $v_\mathrm{max}$ is the maximum projected rotational velocity, $i$ is the disc inclination, $V_\mathrm{asym}$ is the inclination-corrected asymptotic rotational velocity (from Tables~\ref{tab : w-kinemetry} and \ref{tab : t-kinemetry}), and $R$ is the deconvolved half-light radius measured in CO J:3--2 and CO J:7--6.
            
            The emission-line luminosities, $L'_{\mathrm{CO}}$ (in units of K\,km\,s$^{-1}$\,pc$^2$), were calculated using the standard relation \citep{Solomon1992}:
            \begin{equation}
                L'_{\mathrm{CO}} = 3.25 \times 10^7 \times S_{\mathrm{CO}} \Delta v \frac{D_L^2}{(1+z)^3 \nu_{\mathrm{obs}}^2},
            \end{equation}
            where $S_{\mathrm{CO}} \Delta v$ is the line flux in Jy\,km\,s$^{-1}$, $D_L$ is the luminosity distance in megaparsecs, $z$ is the redshift, and $\nu_{\mathrm{obs}}$ is the observed frequency in gigahertz.
            
            The molecular-gas mass is typically estimated from the CO J:1--0 (hereafter CO) luminosity \citep{Solomon2005} using
            \begin{equation}
                M_{\mathrm{mol}} = \alpha_{\mathrm{CO}} \times L'_{\mathrm{CO}},
            \end{equation}
            where $M_{\mathrm{mol}}$ has units of $M_\odot$ and $\alpha_{\mathrm{CO}}$ has units of $M_\odot$ (K\,km\,s$^{-1}$\,pc$^2$)$^{-1}$. There remains a significant debate about the appropriate value of $\alpha_{\mathrm{CO}}$. While for nearby `normal' star-forming galaxies a value of $\alpha_{\mathrm{CO}} \sim 4.3$ $M_\odot$ (K\,km\,s$^{-1}$\,pc$^2$)$^{-1}$ is recommended \citep{Bolatto2013}, significantly lower values are often adopted for starburst galaxies.

            In a seminal analysis of CO radiative transfer and gas dynamics in the starburst nuclei of low-redshift ULIRGs on scales $<$1\,kpc, \citet{DownesAndSolomon1998} found a characteristic value of $\alpha_{\mathrm{CO}} \sim 0.8$ $M_\odot$ (K\,km\,s$^{-1}$\,pc$^2$)$^{-1}$ for H$_2$ + He in such systems. This value, which implies more CO emission per unit molecular-gas mass, is commonly adopted for IR-luminous starbursts, where the gas is not confined to virialised individual clouds \citep{Bolatto2013}.
            
            Given the extreme starburst nature of the four galaxies in \src, and for consistency with I13 and other recent studies of the resolved SK relation, we adopt $\alpha_{\mathrm{CO}} \sim 0.8$ in this work, while noting that a higher value may be more appropriate. More recently, \citet{Dunne2022} argued for a near-universal average value of $\alpha_{\mathrm{CO}}$ in metal-rich galaxies, spanning from main-sequence galaxies to ULIRGs and SMGs. They suggest a value of $\alpha_{\mathrm{CO}} \sim 2.9$ (not including helium), derived for IR-luminous galaxies and SMGs. We later discuss the implications of using this higher value of $\alpha_{\mathrm{CO}}$.
                        
            In Table~\ref{tab : global}, we list the galaxy-integrated molecular-gas masses obtained using the CO J:1--0 luminosities from I13. In Sects.~\ref{sec : resolved line ratios} and \ref{sec : S-K relationship}, we use our resolved CO J:7--6 maps, together with the global CO J:7--6 to CO J:1--0 ratios for each galaxy, to derive resolved molecular-gas masses.
                        
            The neutral-carbon gas mass, and its extrapolation to total molecular mass, has been estimated following \citet{Dunne2022}, under the assumption that the \cishort\ line is optically thin. We emphasise that these authors caution against the use of the \cishort\ line, since its partition function is highly dependent on both temperature and density. As we have observed only the \cishort\ line (and not the [C\,{\sc i}] $^3P_1 \rightarrow\ ^3P_0$ transition), we must assume an excitation temperature ($T_{\rm ex}$) for C\,{\sc i}. When both [C\,{\sc i}] lines have been observed, the derived $T_{\rm ex}$ is typically $\lesssim T_{\rm dust}$ \citep{Stutzki1997, WeiB2005, Popping2017}.
                        
            Table~\ref{tab : global} thus lists the estimated \cishort\ masses for two assumed values of $T_{\rm ex}$, which bracket the estimated dust temperatures from I13 for galaxies W and T. We further assume a gas density of $\log n = 3.5$, typical of SMGs. For this density, Fig.~D1 of \citet{Dunne2022} implies $Q_{21}$ values of 0.17 and 0.23 for $T = 30$\,K and 40\,K, respectively. The [C\,{\sc i}] mass was then calculated using the equivalent of Eq.~8 in \citet{Dunne2022}. The extrapolated molecular-gas mass from this [C\,{\sc i}] mass was estimated using $X_{\rm CI} = 2 \times 10^5$ (see Figs. 6 and 7 of \citealt{Dunne2022}), for the case of $T = 40$\,K.

        \subsection{Continuum and emission-line maps}
            \label{sec: Continuum and Emission-Line Maps}

            Continuum emission at all observed frequencies — corresponding to rest-frame frequencies of 341\,GHz (880\,\micron), 750\,GHz (400\,\micron), 808\,GHz (370\,\micron), and 1160\,GHz (260\,\micron) at $z = 2.41$ — is strongly detected and resolved in galaxies W and T, and relatively weakly detected in galaxies C and M. Among these, the observed 340\,GHz continuum (corresponding to rest-frame 1160\,GHz or 260\,\micron) yields both the highest signal-to-noise ratio and the best spatial resolution ($\sim$0\farcs3; see Fig.~\ref{fig : 1160ghz fluxes}). In the following sections we use the 340\,GHz maps to constrain the global and resolved star-formation rates in each galaxy.
            
\begin{figure*}
                \centering
                \hspace*{-1.28cm} \includegraphics[scale=0.752]{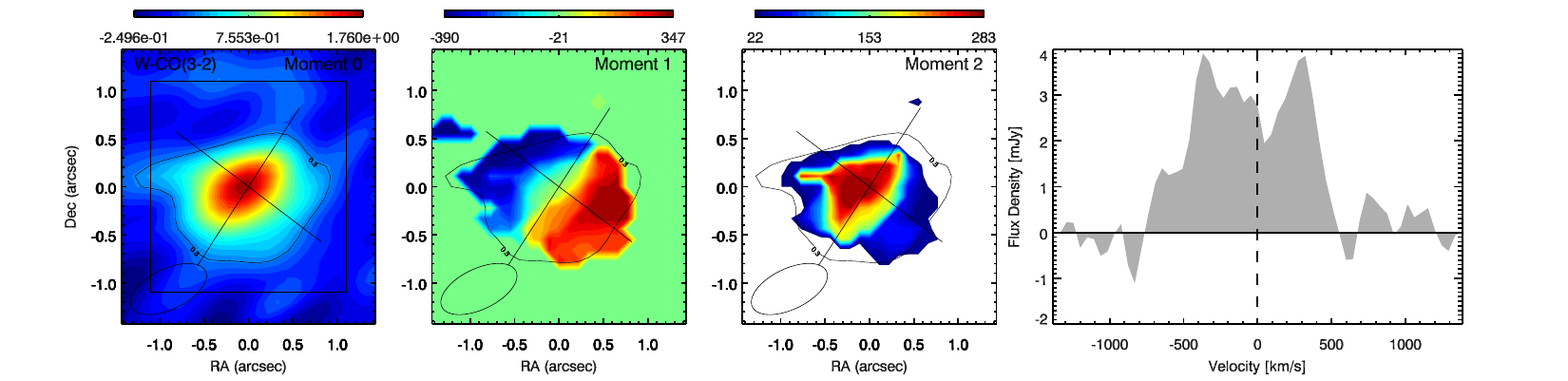} \\
                \hspace*{-1.28cm} \includegraphics[scale=0.752]{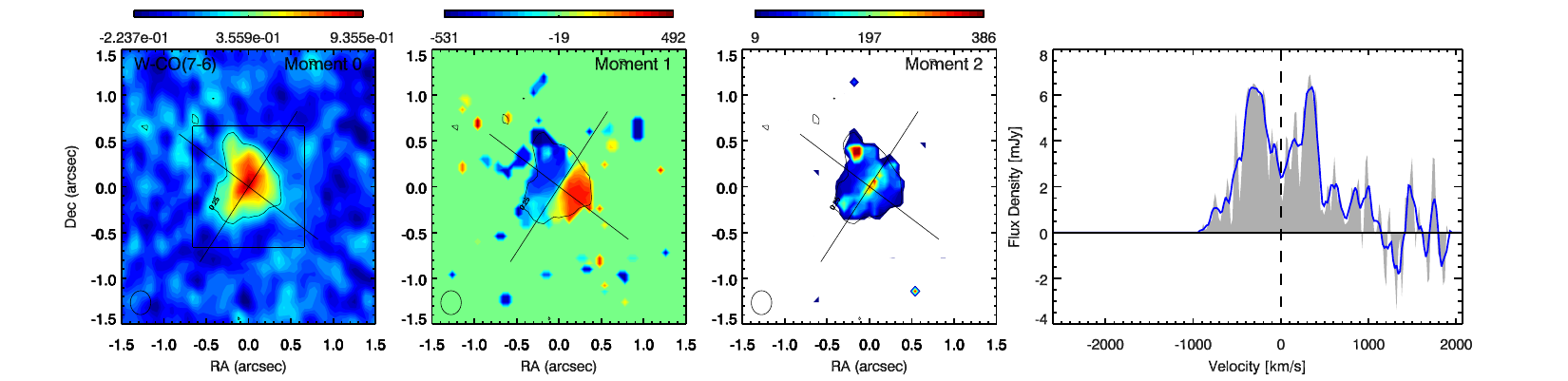} \\
                \hspace*{-1.28cm} \includegraphics[scale=0.752]{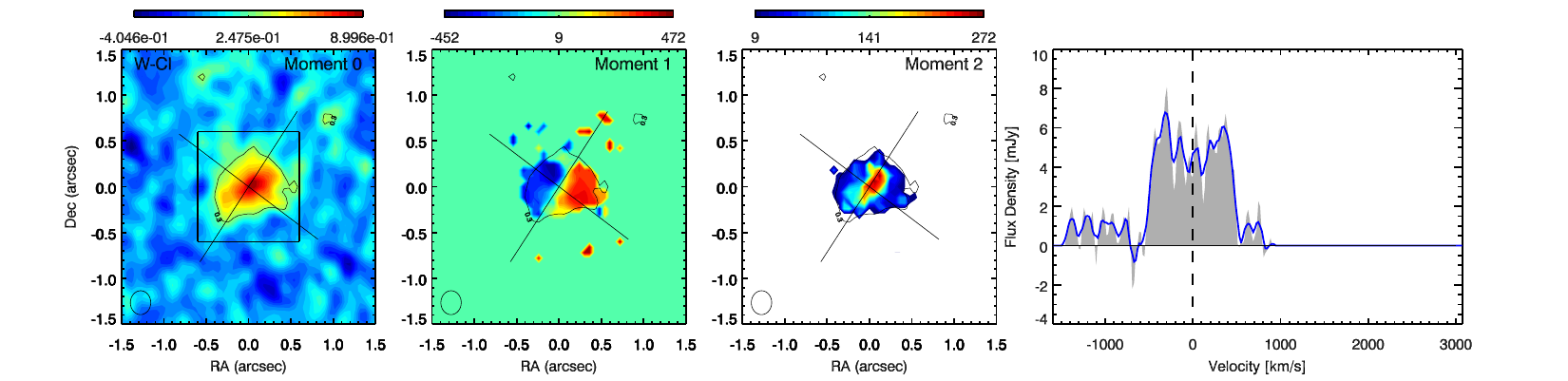}      \\
                \hspace*{-1.28cm} \includegraphics[scale=0.752]{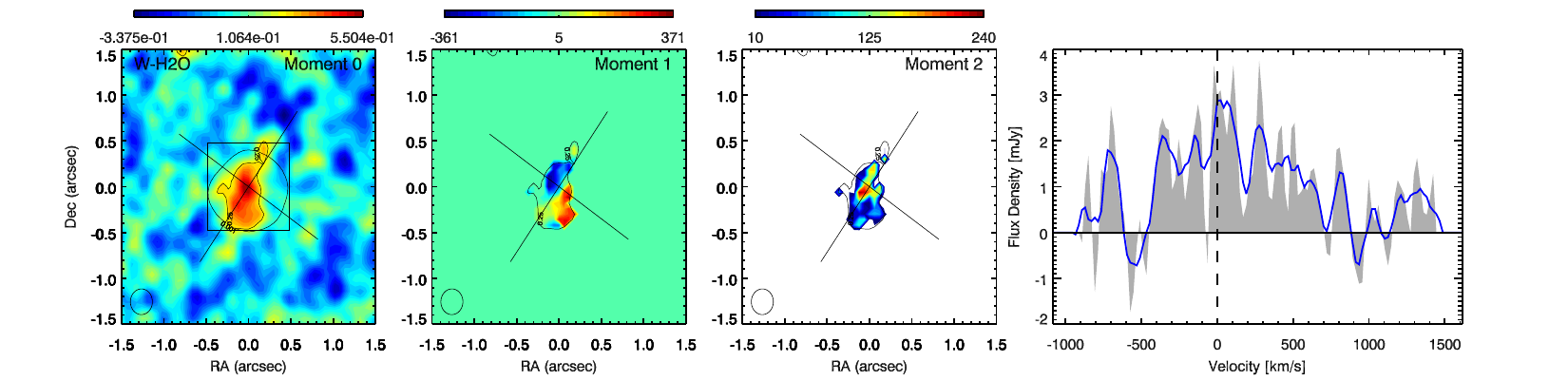}     \\
                \caption{
                Galaxy W. From left to right: Maps of the integrated flux (moment 0; units of Jy \kms\ beam$^{-1}$), intensity-weighted average velocity (moment 1; \kms\ relative to systemic), velocity dispersion (moment 2; \kms), and the galaxy-integrated line profile, of the detected emission lines. From top to bottom the lines are:  CO J:3--2, CO J:7--6, \cifull\ , and \h2ofull. In the three left-most columns, the colours follow the respective colour bars, the synthesised beam is shown at the lower left, and axis units are arcseconds with the same central position used in all panels. For ease, the kinematic major and minor axes are shown in solid black lines and a single specific flux contour from the moment 0 image is shown in all panels of the same row. All moment maps were made from natural-weighted data cubes. The right-most column shows the corresponding galaxy-integrated line profile extracted within the square aperture shown in the corresponding left-most panel. The line profiles are shown both at the observed spectral resolution (grey histograms; spectral resolutions of -- top to bottom -- 46 km s$^{-1}$, 19.7 km s$^{-1}$, 19.7 km s$^{-1}$, and 21.2 km s$^{-1}$, per channel), and at a smoothed resolution (blue solid lines in the lower three panels) of $\sim$ 100 \kms. The line profiles of CO J:7--6 and \cifull\ have been de-blended as is explained in Sect.~\ref{sec: Galaxy-integrated spectral properties}.
                }
                \label{fig : w-moments & profiles}
            \end{figure*}

\begin{figure*}
                \centering
                \hspace*{-1.28cm} \includegraphics[scale=0.753]{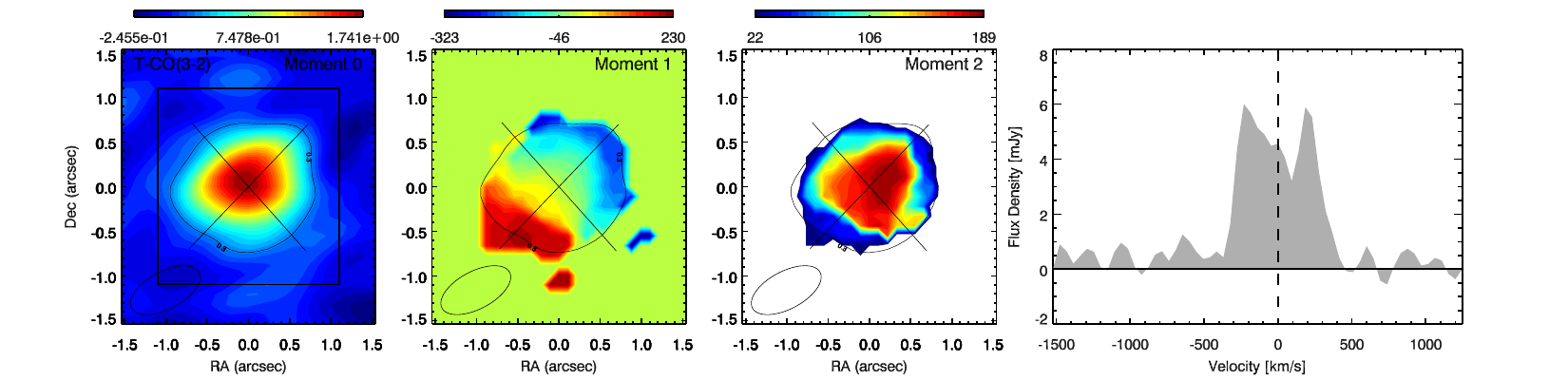} \\
                \hspace*{-1.28cm} \includegraphics[scale=0.753]{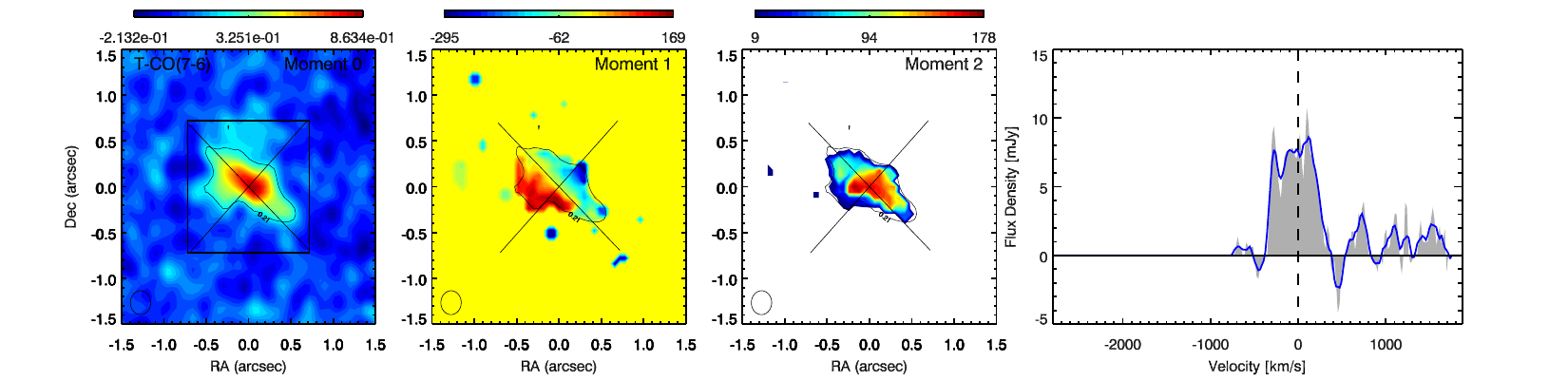} \\
                \hspace*{-1.28cm} \includegraphics[scale=0.753]{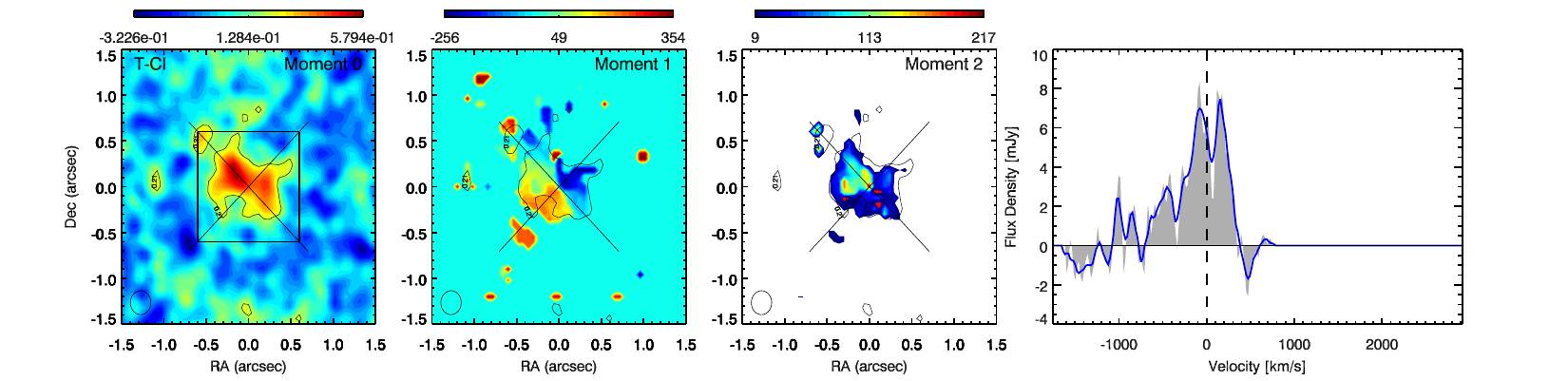}      \\
                \hspace*{-1.28cm} \includegraphics[scale=0.753]{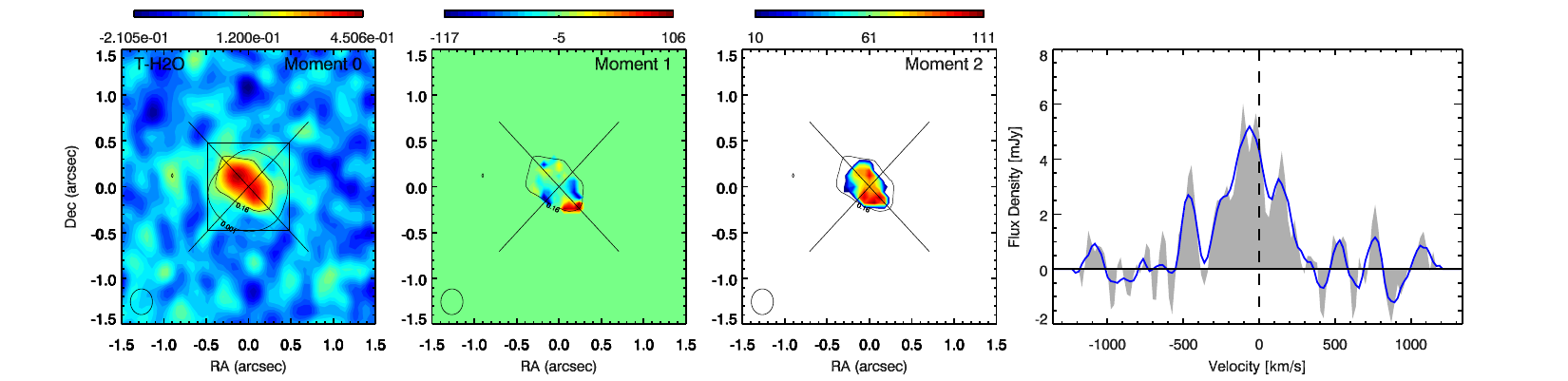}     \\
                \caption{
                Same as Fig.~\ref{fig : w-moments & profiles} but for galaxy T. Here the relatively narrow line profiles, as compared to W, allow for a cleaner separation of the galaxy-wide CO J:7--6 and \cifull\ line profiles.
                }
                \label{fig : t-moments & profiles}
            \end{figure*}

\begin{figure*}
                \centering
                \hspace*{-1.28cm} \includegraphics[scale=0.753]{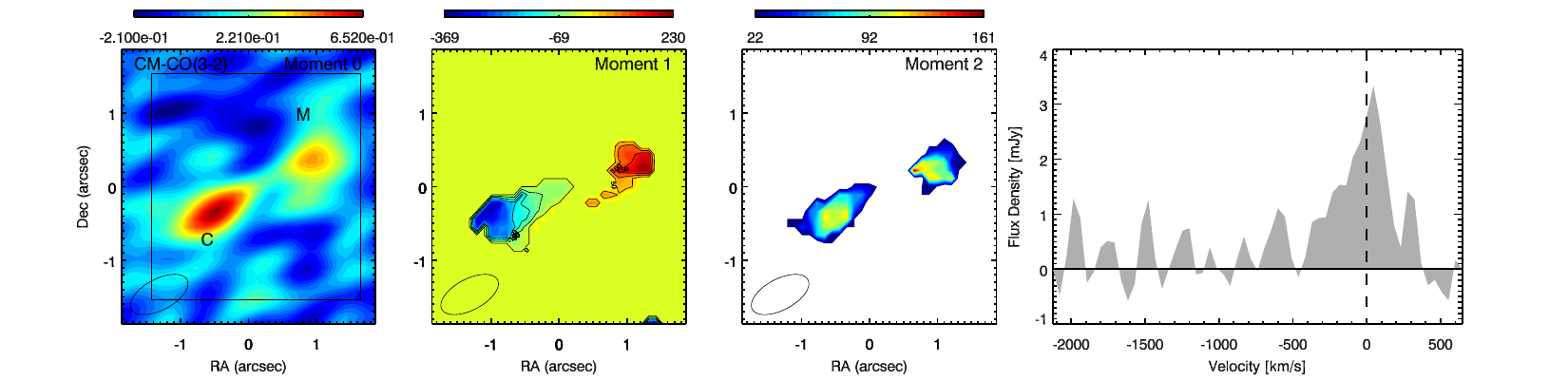} \\
                \hspace*{-1.28cm} \includegraphics[scale=0.753]{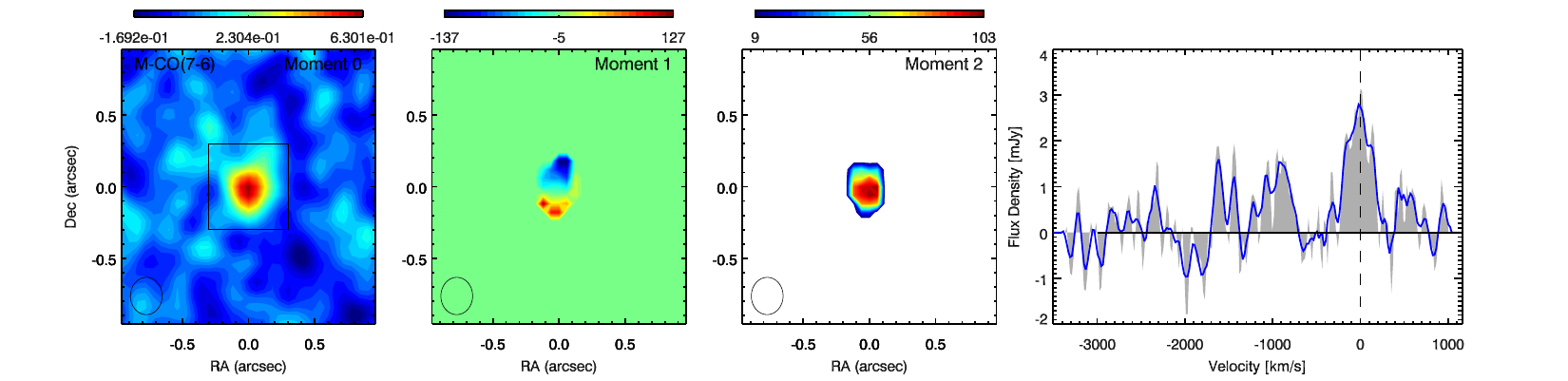}  \\
                \hspace*{-1.28cm} \includegraphics[scale=0.753]{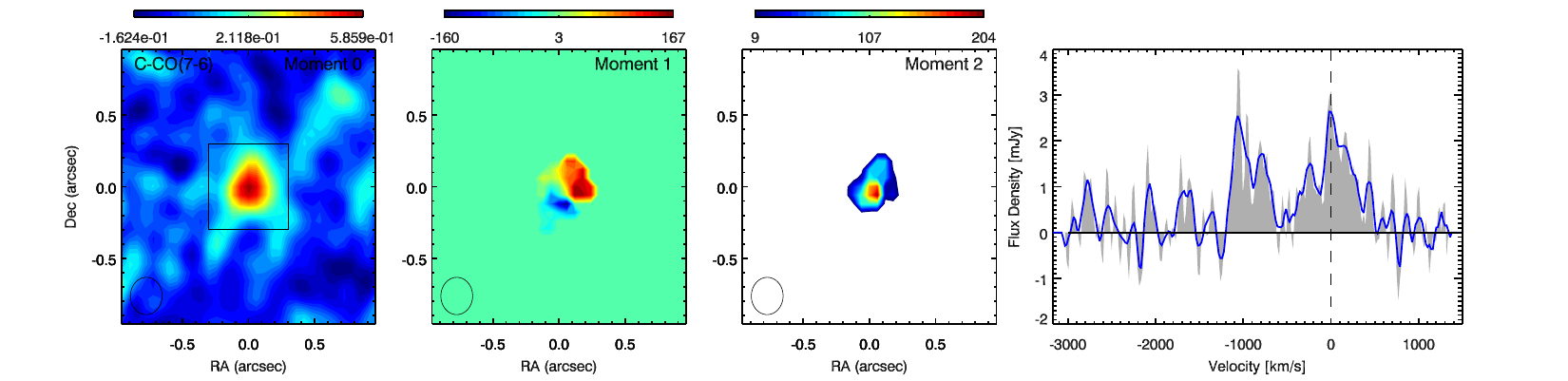}  \\
                \caption{
                Similar to Fig.~\ref{fig : w-moments & profiles} but for  galaxies M and C. The top row shows the results for the CO J:3--2 line in both M and C together. The middle and bottom rows show the results for the CO J:7--6 line separately for M and C, respectively. In the right-most panels of the middle and bottom rows  (the galaxy-integrated profiles of the CO J:7--6 lines in M and C, respectively) the CO J:7--6 line covers a velocity range of approximately $\pm 300$ \kms\ and 500 \kms, respectively; for both galaxies the \cifull\ line is also clearly visible at lower ($\sim$1000 \kms\ to the blue) velocities. We do not show the equivalent moment maps for the \cifull\ line in M and C as they are significantly noisier.
                }
                \label{fig : cm-moments & profiles}
            \end{figure*}

            Moment maps (0 = integrated flux, 1 = intensity-weighted velocity map, 2 = velocity-dispersion map) of the CO J:3--2, CO J:7--6, \cishort, and H$_2$O emission lines are shown in Figs.~\ref{fig : w-moments & profiles}, \ref{fig : t-moments & profiles}, and \ref{fig : cm-moments & profiles}, for galaxies W, T, and C and M, respectively. The galaxy-integrated line profiles shown in the rightmost panels of these figures were extracted from the square apertures indicated in the leftmost (moment 0) maps.
            
            For galaxy W, all lines were extracted within a $3\arcsec \times 3\arcsec$ aperture centred at RA = $08^h49^m33\fs592$, Dec = $+02\degr14\arcmin44\farcs618$. For galaxy T, a similar $3\arcsec \times 3\arcsec$ aperture was used, centred at RA = $08^h49^m32\fs960$, Dec = $+02\degr14\arcmin39\farcs697$.
                        
            The CO J:3--2 line profile for galaxies C and M (combined) was extracted using a $3\farcs8 \times 3\farcs8$ aperture centred at RA = $08^h49^m33\fs868$, Dec = $+02\degr14\arcmin45\farcs219$. In contrast, the individual CO J:7--6 line profiles of galaxies C and M were extracted from smaller $0\farcs5 \times 0\farcs5$ apertures centred on each galaxy.

    \section{Results}
        \label{sec : results}
        
        In this section we present the main observational results derived from the data-analysis methods described in Sect.~\ref{sec : analysis methods}. We focus on the integrated line profiles and ratios, the spatial extent and morphology of the continuum and line emission, the resolved kinematics of the molecular and atomic gas, a comparison with rotation-only models using position–velocity (PV) diagrams, and finally other galaxy-resolved quantities including resolved emission-line ratios and a resolved SK relation. Unless stated otherwise, the discussion is centred on the four main components of HATLAS\,J084933: galaxies W, T, M, and C.

        \subsection{Galaxy-integrated spectral properties}
            \label{sec: Galaxy-integrated spectral properties}
        
            \begin{figure}[ht]
                \centering
                \includegraphics[scale=0.87]{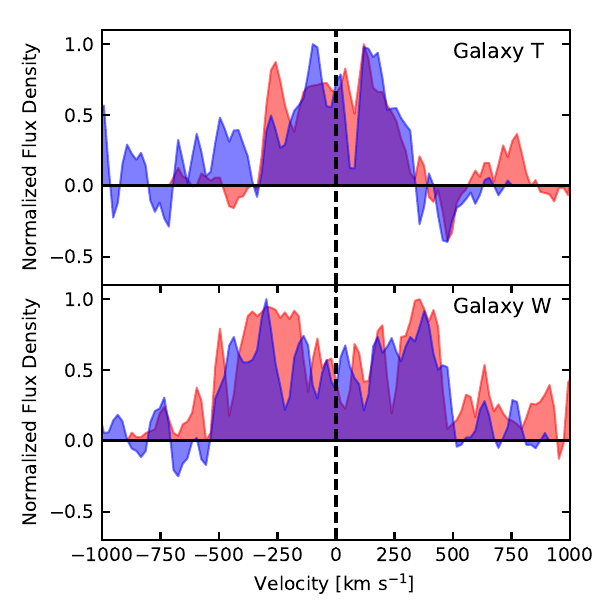}
                \caption{Comparison of the galaxy-integrated normalised spectral profiles of the  CO J:7--6 (red) and [C\,{\sc i}] 2--1 (light blue) lines for galaxies T (top panel) and W (bottom panel). Overlap regions of the two spectra appear darker blue. The dashed black line shows our adopted zero velocity.
                }
                \label{fig : comparison of profiles}
            \end{figure}

            A direct comparison of the galaxy-integrated line profiles of CO J:7--6 and \cishort\ in galaxies W and T is shown in Fig.~\ref{fig : comparison of profiles}. These profiles were separated using the velocity-channel method based on the rotation-only models developed in Sect.~\ref{sec: line separation method}, which enables the disentangling of blended emission from both lines in the data cube. In galaxy W, both profiles exhibit a double-horned structure, typically attributed to significant emission from gas located at radii where the rotation curve is flat \citep{deBlokAndWalter2014}. Additionally, while the redshifted emission from both lines is relatively well matched (in a normalised sense), the CO J:7--6 line shows a notable excess over \cishort\ at blueshifted velocities. Both features, although less pronounced, can also be seen in galaxy T.

            \begin{figure}
                \hspace{-0.5cm}
                \includegraphics[scale=0.9]{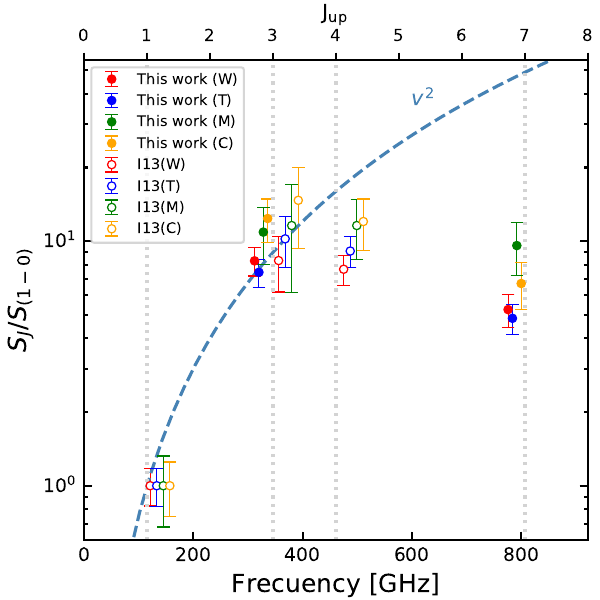} \\
                \caption{CO ladder in all four galaxies in \src: W (red), T (blue), M (green), and C (yellow). The $x$ axis is labelled with the rotational quantum number of the upper state (top; 1, 3, 4, and 7) and the rest frequency of the line (bottom). Small  displacements in the $x$ axis are used for the different galaxies to avoid confusion, and vertical lines demarcate the true $x$ positions of the points. The $y$ axis shows the line flux normalised to that of CO J:1--0. Line fluxes from this work are shown with solid circles and those from I13 are shown with circles. The dashed line shows the CO ladder expectation for the case of a constant brightness temperature on the Rayleigh-Jeans (RJ) scale, i.e.\ $S \sim \nu^2$ (note that the RJ approximation is not valid for high $J$).
                }
                \label{fig : co ladder}
            \end{figure}

            The variation in galaxy-integrated CO line luminosity with rotational transition — the so-called `CO ladder' — for each of the four galaxies in \src\ is shown in Fig.~\ref{fig : co ladder}. Galaxy T appears to peak at CO J:4--3, while W, C, and M display flatter distributions between CO J:3--2 and CO J:4--3.

            Note that the CO J:4--3 data from I13 were obtained with IRAM PdBI (six antennas) in its most extended configuration (6Aq). The relatively limited $uv$ coverage of that set-up (136–760~m, i.e. $\sim$60–340~k$\lambda$) compared to our ALMA CO J:3--2 observations (15–1467~m, i.e. $\sim$5–500~k$\lambda$) suggests that the former images may have resolved out part of the extended emission. Conservatively, we conclude that the CO ladder peaks somewhere between CO J:3--2 and CO J:6--5 in all four galaxies.
            
            A peak around J~$\sim$5 is consistent with what is typically observed in SMGs \citep{CarilliAndWalter2013}. We emphasise that the uncertainty in the exact location of the peak does not introduce additional uncertainty into our later conversion between CO J:7--6 and CO J:1--0 luminosities.
            
        \subsection{Resolved continuum morphology and spectral slope variations}
            \label{sec : resolved continuum}
    
            \begin{figure*}[ht]
                \includegraphics[scale=0.55]{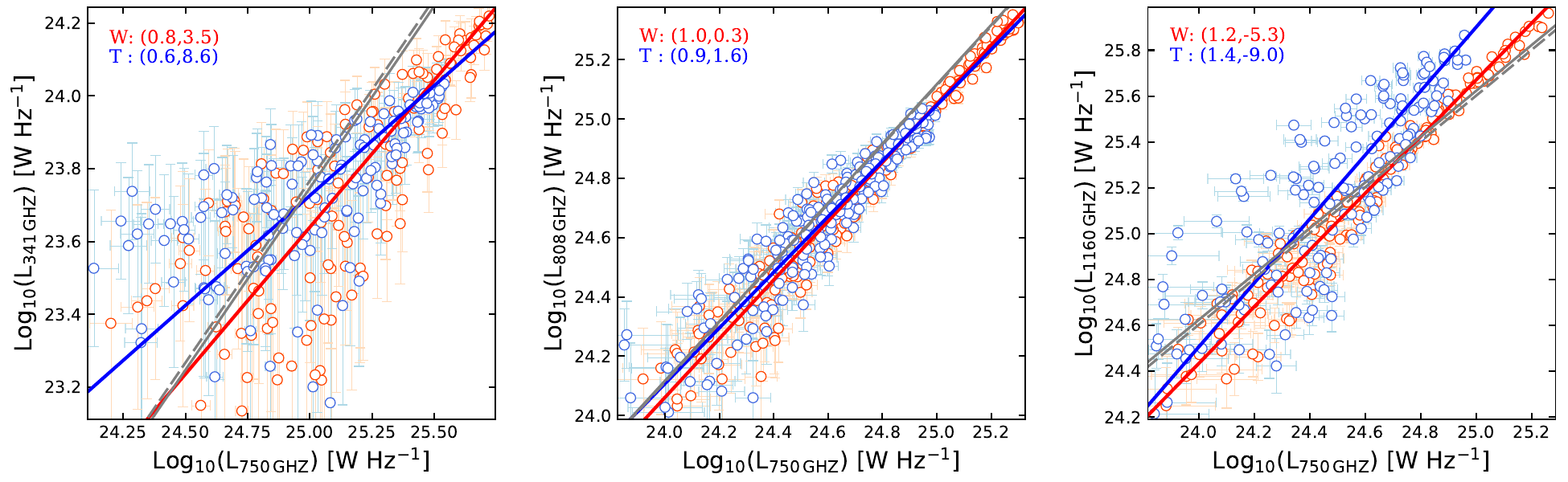}
                \caption{Relationships between the (resolved) continuum luminosities at rest-frame 341~GHz (880\micron), 750~GHz (400\micron), 808~GHz (370\micron), and 1160~GHz (260\micron) for galaxies W (red circles) and T (blue circles). The two maps used for each panel were convolved to a common resolution. Each data point was calculated over an aperture equivalent to the FWHM of the synthesised beam, with apertures spaced by half a synthesised beam; i.e.\ roughly a quarter of the data points are independent measurements. Error bars for each point are shown in light red and light blue for galaxies W and T, respectively. The solid lines in the corresponding colour delineate the linear fits to the data; the coefficients of these fits (slope, intercept) are listed in the panel in the corresponding colour. The grey lines show the expected relationships for greybody emission (with $\beta=2.0$) for dust at temperature, $T_{\rm dust}=40$~K (solid) and 36~K (dashed), the estimated dust temperatures derived by I13 for W and T, respectively. 
                }
                \label{fig : rel. of continuum luminosities}
            \end{figure*}

            We explore the resolved continuum spectral slopes of galaxies W and T by comparing the four observed continuum luminosities across multiple apertures (Fig.~\ref{fig : rel. of continuum luminosities}). To generate the data in each panel, the higher-resolution image was convolved with a Gaussian to match the resolution of the lower-resolution image. Fluxes were then extracted from apertures with sizes equal to the FWHM of the synthesised beam of the lower-resolution image, spaced by half a beam width. Consequently, only about a quarter of the data points are statistically independent.

            I13 fitted the global sub-mm SEDs of galaxies W and T with a single-temperature greybody model (assuming $\beta = 2.0$), yielding dust temperatures of $39.8 \pm 0.1$~K and $36.1 \pm 1.1$~K, respectively. The predictions of greybody emission for these parameters are shown in all panels of Fig.~\ref{fig : rel. of continuum luminosities}. Between 370 and 400\,\micron\ (middle panel), the resolved flux ratios are consistent with the globally derived temperatures.

            Galaxy T shows significant excess continuum emission at both the longest and shortest wavelengths. All apertures in T exhibit a consistent flux excess at 880\,\micron\ (left panel), which may indicate the presence of cooler dust or an additional source of emission at this wavelength. In the nuclear apertures of T, the excess at 260\,\micron\ (right panel) suggests either a significantly higher temperature in a single dust component or the presence of a second, hotter dust component.

            Figure~\ref{fig : flux sizes} compares the spatial distribution of the CO J:7--6, \cishort, and rest-frame 1160\,GHz continuum emission along the major and minor kinematic axes of galaxies W and T. These axes were defined under the assumption of rotation-dominated kinematics (see Sect.~\ref{sec: line separation method}).

            Interestingly, the continuum emission of galaxy T is primarily extended along its kinematic minor axis (thin grey line in Fig.~\ref{fig : flux sizes}), matching the morphology of the CO J:7--6 emission. At the faintest levels, however, T also shows an extension along its major axis, particularly towards the south-east. In contrast, the brighter continuum emission in galaxy W appears extended along a position angle of $\sim$100\degr, intermediate between its major and minor kinematic axes. At lower flux levels, W shows more extended emission along its major axis.

            Galaxy C appears extended along PA $\sim$105\degr, with emission to the west as a connected extension and to the east as a lower-significance, detached knot. The continuum distribution in W and T along the kinematic axes can also be appreciated in Fig.~\ref{fig : flux sizes}, where it is directly compared with the emission-line flux distributions.

            These continuum morphologies and spectral slopes provide important constraints on the spatial distribution of dust-obscured star formation, which we explore further in Sect.~\ref{sec : S-K relationship}.

        \subsection{Resolved emission-line morphologies and kinematics}
            \label{sec : galaxy sizes and emission-line maps}
       
            All detected emission lines — CO J:3--2, CO J:7--6, \cishort, and H$_2$O — are resolved across several synthesised beams in W and T, and the CO J:7--6 and \cishort\ lines are marginally resolved in C and M. Galaxies W and T show the largest spatial extents in both continuum and line emission.
            
\begin{figure}[ht]
                \centering
                \includegraphics[scale=0.57]{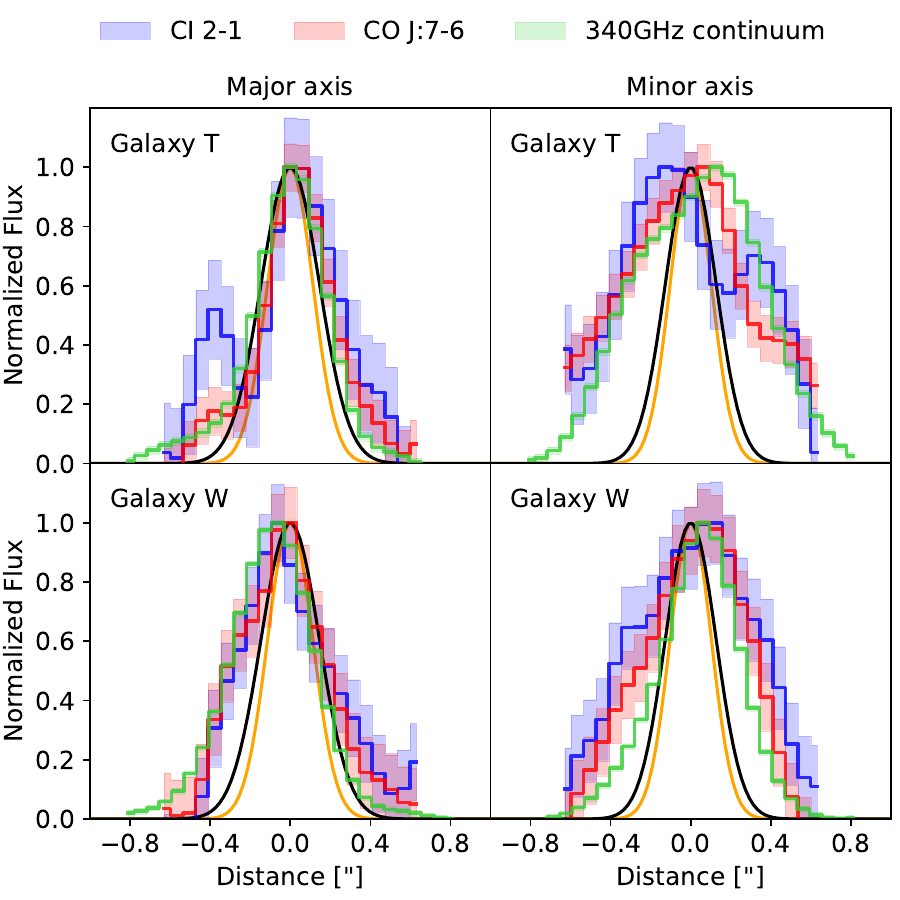}
                \caption{
                Spatial extent of the \co7-6\ and \cishort\ emission lines  and rest-frame 1160-GHz continuum emission in galaxy T (top row) and galaxy W (bottom row) along the kinematic major (PA=135\arcdeg\ for T and PA=55\arcdeg\ for W; left panels) and kinematic minor axes (right panels). Each panel shows the normalised flux, extracted along a one-pixel-wide slit oriented along the corresponding PA, of the observed-frame 340~GHz continuum emission (green), and the \co7-6\  (red) and  \cishort\  (blue) emission lines.  The shaded region with the corresponding colour delineates the 1$\sigma$ noise in the neighbourhood of each extracted pixel. The major axis of the synthesised beam of the \co7-6\ and \cishort\ maps (340-GHz continuum map) is shown by the orange (black) normalised Gaussian;  both synthesised beams are close to circular. Both lines and continuum are clearly resolved in all panels, with similar extensions in both emission lines and the continuum. The deconvolved sizes along these PAs are listed in Tables~\ref{tab : continuum sizes} and \ref{tab : co76 sizes}.
                }
                \label{fig : flux sizes}
            \end{figure}

            In Fig.~\ref{fig : flux sizes}, within the noise, the distributions of the gas and continuum emission are roughly similar in galaxy W, though there is some indication that the gas (especially \cishort) emission is more extended than the continuum emission along the kinematic minor axis. Additionally, and unexpectedly, the emission from both atomic and molecular gas is more extended along the minor axis, rather than the major axis. Even more significant differences are seen in galaxy T. Here the gas and dust are more extended along the (kinematic) minor axis rather than the major axis. Along the minor axis there are significant offsets between the locations of the peak CO J:7--6, \cishort, and continuum emission. The continuum peak (the `SW knot') has a significantly higher CO-to-\cishort\ ratio than the NE extension; similar to \cishort, the H$_2$O emission is also stronger in the NE extension than in the SW knot (see the next section). The continuum-to-gas ratios are highest in the NE extension and lowest in the SW knot. Along the kinematic major axis, the gas and continuum are relatively similarly distributed, except for a \cishort-rich region 0\farcs4 from the nucleus. The deconvolved sizes of W and T along the PAs shown in Fig.~\ref{fig : flux sizes} are listed in Tables~\ref{tab : continuum sizes} and \ref{tab : co76 sizes}.
       
            \begin{table}[ht]
                \caption{Deconvolved sizes of the observed-frame 340 GHz emission.}
                \centering
                \begin{tabular}{c c c}
                \hline
                Galaxy & kinematic             & kinematic             \\
                       & minor-axis            & major-axis            \\
                \hline
                W      & 0.46\arcsec(3.80 kpc) & 0.45\arcsec(3.71 kpc) \\
                T      & 0.75\arcsec(6.19 kpc) & 0.39\arcsec(3.22 kpc) \\
                \hline
                \end{tabular}
                \newline
                {
                \footnotesize{Values listed are the FWHM of the deconvolved Gaussian.}
                }
                \label{tab : continuum sizes}
            \end{table}
        
            \begin{table}[ht]
                \caption{Deconvolved sizes of the CO J:7--6 emission region.}
                \centering
                \begin{tabular}{c c c}
                \hline
                Galaxy & kinematic              & kinematic             \\
                       & minor-axis             & major-axis            \\
                \hline
                W      & 0.58\arcsec(4.79  kpc) & 0.53\arcsec(4.37 kpc) \\
                T      & 0.75\arcsec(6.19 kpc)  & 0.36\arcsec(2.97 kpc) \\
                \hline
                \end{tabular}
                \newline
                {
                \footnotesize{Values listed are the FWHM of the deconvolved Gaussian.}
                }
                \label{tab : co76 sizes}
            \end{table}
    
            All four galaxies show ordered velocity fields. In W and T, both CO and \cishort\ show consistent blueshifted and redshifted sides and similar peak-to-peak velocities. The molecular gas velocity fields are best appreciated in the CO J:7--6 (rather than CO J:3--2) velocity maps, which have both the highest signal-to-noise ratio and the highest spatial resolution. Galaxies W and T show (projected) maximum velocities of roughly $\pm 500$ \kms\ and $\pm 270$ \kms, respectively. Galaxies M and C have smaller velocity gradients, with $v_{\mathrm{max}} = 132$ km s$^{-1}$ (264 km s$^{-1}$ peak-to-peak) and $v_{\mathrm{max}} = 164$ km s$^{-1}$ (327 km s$^{-1}$ peak-to-peak), respectively. Note that the internal kinematics of C and M are resolved for the first time in our CO J:7--6 and \cishort\ maps. Further, the relatively low-spatial-resolution CO J:3--2 maps show an almost continuous (in flux and velocity) bridge between C and M. The CO and \cishort\ dispersion maps show higher dispersions near the nucleus and along the kinematic minor axis; the former can be partly due to the rotational velocity gradient along the major axis. The observed velocity fields are best interpreted as ordered rotation (Sect.~\ref{sec: Rotational Model Construction}) in a galaxy disc.

            While the CO and \cishort\ lines follow roughly similar kinematics in W, there are notable differences in both their morphology and kinematics. For example,\ CO J:7--6 shows a prominent, blueshifted, highly dispersive north-east (NE) clump that is not present in \cishort. These differences are the primary reason for the different global emission-line profiles presented in the previous section.
            
            In galaxy T, the CO and \cishort\ emission lines show relatively similar kinematics. While CO J:7--6 emission is roughly centred on the `nucleus', the \cishort\ and \water\ line emission straddle the nucleus, similar to what is seen in the (rest-frame) 1160~GHz continuum emission. Note that galaxy T is more extended along its kinematic minor axis than its major axis (see the middle panel of Fig.~\ref{fig : t-moments & profiles}); this was not obvious in the earlier I13 maps due to their lower resolution. The extension along the minor axis is best appreciated in Fig.~\ref{fig : flux sizes} and Tables~\ref{tab : continuum sizes} and \ref{tab : co76 sizes}. This galaxy is already known to have a $2\times$ flux magnification due to lensing (I13). If the velocity field in T is interpreted as rotation, then the ratio of the minor-to-major axes in this galaxy implies that the lensing-produced spatial magnification is $\gtrsim 2\times$ along the direction of the minor axis.

            Both C and M are also resolved kinematically in the CO J:7--6 line (Fig.~\ref{fig : cm-moments & profiles}) and the \cishort\ line (not shown): if their velocity fields are due to rotating gas, then the major-axis PAs of C and M are $\sim -$30\arcdeg\ and $\sim$160\arcdeg, respectively. The lower-resolution CO J:3--2 maps appear to show a continuous bridge, in integrated flux and in velocity, between the two galaxies. Given their small spatial and velocity separation, it is likely that these two galaxies are interacting — a not uncommon feature of SMGs (e.g.\ \citealt{Ivison2007, Engel2010}).
            
            The galaxy-integrated CO and \cishort\ emission-line profiles in W and T can be interpreted as double-peaked, or as having a central plateau, which indicates that the molecular gas extends well into the part of the disc with a flat rotation curve.
    
        \subsection{Rotation-only model comparison and PV evidence of gas motions}
            \label{sec : rotation model}
    
            \begin{figure*}[ht]
                \centering
                \includegraphics[scale=0.65]{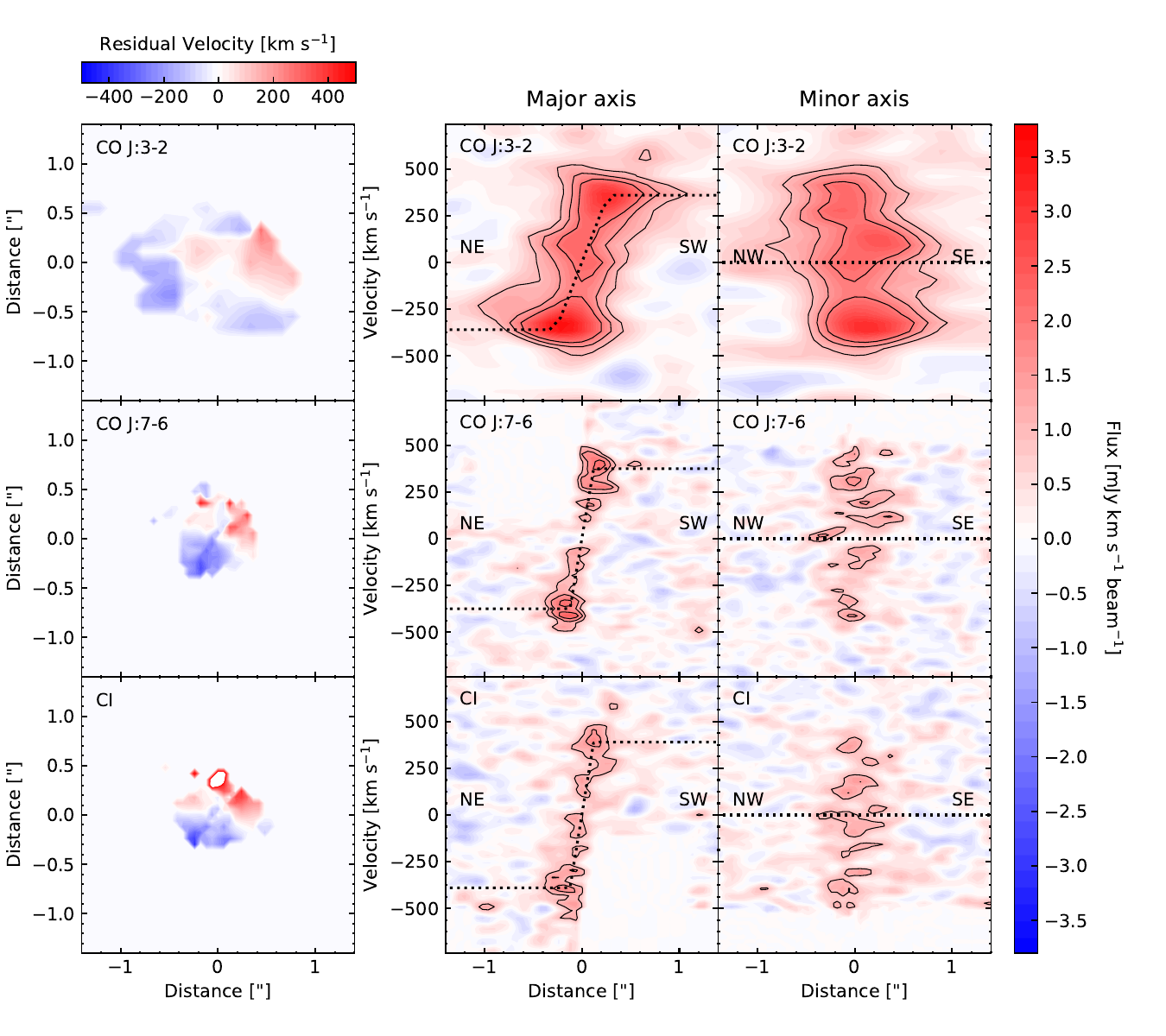}
                \caption{
                Comparison of observed velocity fields to the \kin-derived rotation-only models (see Sect. 3.2) for galaxy W. Top row: Velocity field residuals (observed velocity minus our rotation-only model), and the position-velocity diagrams along the major (middle) and minor (right) axes of W, for the CO J:3-2 line. In the middle and right panels, emission line fluxes are shown in colour following the common colour bar to the far right and in black contours (1, 1.5 and 2 mJy \kms $\mathrm{beam^{-1}}$). The dashed black lines in these panels show the predictions of our adopted rotation model (see Sect. \ref{sec : rotation model}). The middle and bottom rows show the equivalent plots for the CO J:7-6 and \cishort\ lines, respectively. All three left panels follow the colour bar on the top of the left panels, and areas with low signal to noise in the moment 0 map are masked . All other panels follow the colour bar shown on the right of the figure. At our adopted distance for HATLAS\,J084933, $1''$ corresponds to 8.25 kpc. Each observed velocity field was input to the \kin\ package to determine the kinematic PA and inclination (see text): these independently derived values are similar for all lines except CO J:3--2. The pure rotation models were then constructed (by eye) using these values of PA and inclination together with a simple model of solid body rotation which changes to a flat rotation beyond a certain radius.
                }
                \label{fig : w-pv diagrams}
            \end{figure*}
        
            \begin{figure*}[ht]
                \centering 
                \includegraphics[scale=0.65]{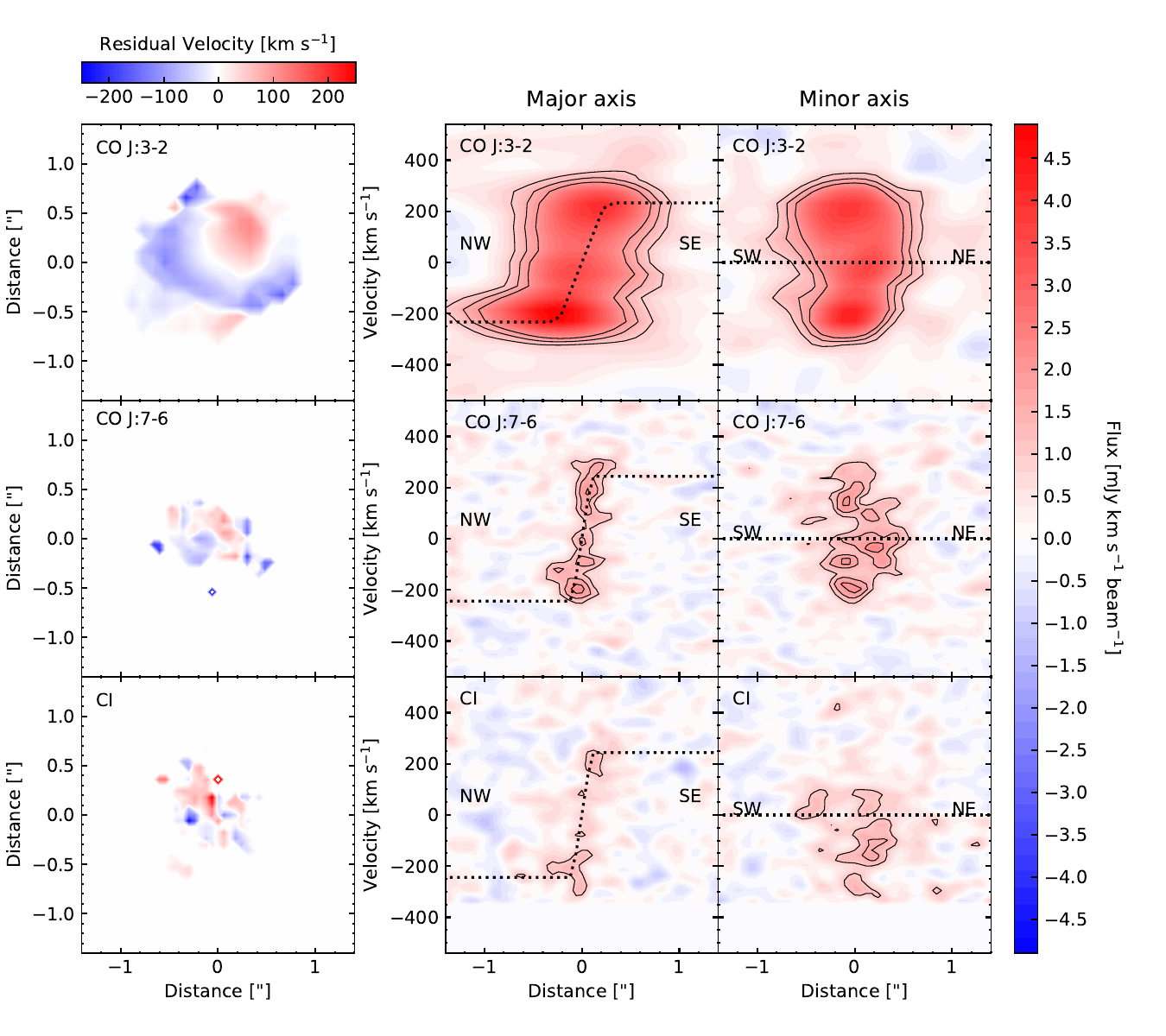}
                \caption{
                Same as Fig.~\ref{fig : w-pv diagrams} but for galaxy T of  HATLAS J084933.4+021443.
                }
                \label{fig : t-pv diagrams}
            \end{figure*}
    
            Results of the rotation-only modelling of galaxies W and T are shown in Figs.~\ref{fig : w-pv diagrams} and \ref{fig : t-pv diagrams}, respectively. For both W and T, the \kin\ analysis, and thus the rotation-only models, are very similar for both CO J:7--6 and \cishort. The models for CO J:3--2 also give similar PAs and inclinations, but the best-fit rotation curve is relatively smooth, as expected given the lower spatial resolution of the data. The left-hand columns of Figs.~\ref{fig : w-pv diagrams} and \ref{fig : t-pv diagrams} present the velocity residuals, i.e.\ $V_{\mathrm{observed}} - V_{\mathrm{model}}$, in order to highlight deviations from pure rotation. In W, we consistently see red velocities to the north-west (NW) and blue velocities to the south-east (SE), which could imply outflows in the disc plane. The residuals of T show no clear signs of non-circular rotation.
    
            \begin{figure}
                \includegraphics[scale=0.6]{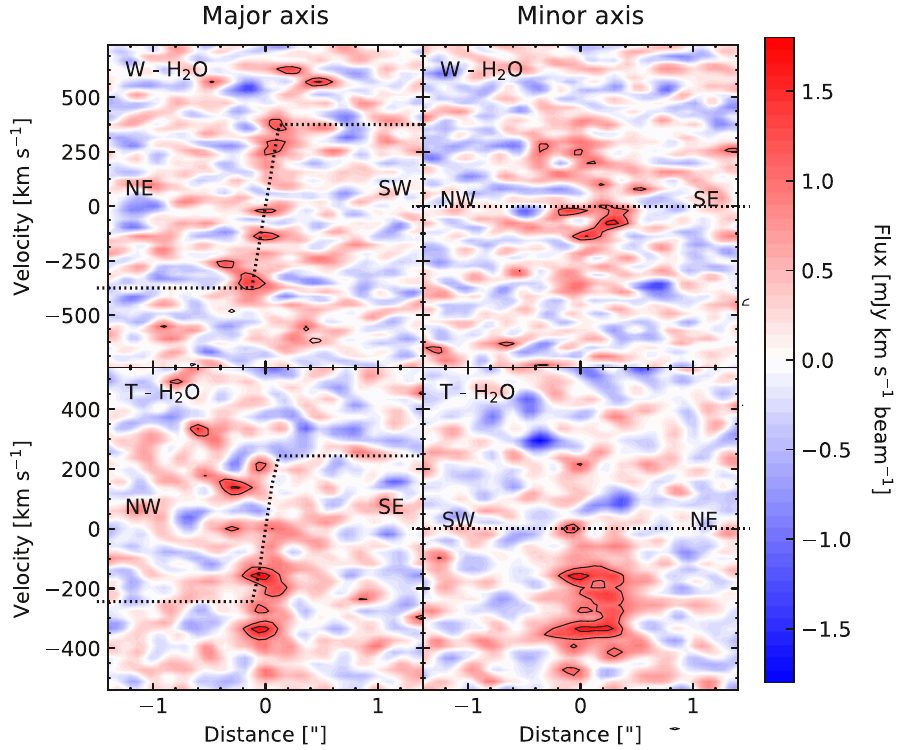}
                \caption{
                Position-velocity (PV) diagrams for \water\ emission along the major (left) and minor (right) axes in W (top) and T (bottom).
                }
                \label{fig : wt-pv diagrams}
            \end{figure}
       
            Position–velocity (PV) diagrams of the CO and \cishort\ lines are shown in Figs.~\ref{fig : w-pv diagrams} and \ref{fig : t-pv diagrams} for galaxies W and T, respectively, and PV diagrams for the \water\ line in W and T are shown in Fig.~\ref{fig : wt-pv diagrams}. The PV diagrams were extracted along the kinematic major and minor axes discussed in the previous section: kinematic major-axis PAs of 55\arcdeg\ and 135\arcdeg\ for W and T, respectively (see Figs.~\ref{fig : w-pv diagrams} and \ref{fig : t-pv diagrams}), using a pseudo-slit of width 0\farcs06 for CO J:7--6 and [C\,{\sc i}] 2--1, and 0\farcs11 for CO J:3--2. They are thus limited in spatial resolution by the intrinsic resolution of the images. On each PV diagram, we have overlaid the predictions of our toy rotation model (dashed black line) described in Sect.~\ref{sec : rotation model}. Along the major axis, in both W and T, the toy rotation model follows the velocity structure of the CO and \cishort\ gas well: in fact, especially in CO J:3--2, one can discern the point at which the presumed rotation changes from solid body to flat. The gas velocities do show interesting wiggles away from the predictions of the rotation model—for example, at radii close to the nucleus and to the SW in the CO J:3--2 major-axis PV diagram—but higher spatial resolution and signal-to-noise is required to model these deviations. The kinematic signatures of W and T in their PV diagrams along the posited kinematic major axis, specifically the velocity gradient seen close to the nucleus, provide strong evidence that the kinematics are rotation-dominated. Outflows within the disc would produce a more abrupt change in velocity from one side of the disc to the other. Along the kinematic minor axes of W and T, the data are more difficult to interpret: we expect to see zero velocities at all offsets, but smearing from the relatively large synthesised beam means that multiple velocity components are seen at all spatial offsets along the minor axis. This is unfortunate, since any outflow within the plane of the disc would be most apparent along the minor axis. In W, the minor-axis PV diagram of the CO J:3--2 line shows a $\sim$100~\kms\ feature to the SE. In fact, this distortion in the zero-velocity line is also clearly seen in the equivalent velocity field (e.g.\ top row of Fig.~\ref{fig : w-pv diagrams}). The equivalent, but fainter, blueshifted component is seen to the NW. This could trace a molecular outflow in the disc if the SE is the far side of the disc. However, many alternative explanations are possible, including bar-driven distortions in the gas kinematics, and thus we cannot at this point present evidence for outflows or inflows.

            The PV diagrams of the \water\ line in W and T are presented in Fig.~\ref{fig : wt-pv diagrams}. These are relatively noisy, but appear to follow the kinematics seen in the other lines. In T, the \water\ line is brightest to the NW.
            
            We tested for P~Cygni or simple absorption-line profiles at similar velocity offsets (from systemic) in all of \co7-6, \cishort, and \water\ by modelling the galaxy-wide spectra as a sum of multiple Gaussian emission- and absorption-line components. The velocities and widths were kept consistent across all species, while only the amplitudes of the components were allowed to vary. The only potential absorption line seen in both \co7-6 and \cishort\ is one at $450$~km\,s$^{-1}$, with a FWHM of $120$~km\,s$^{-1}$. However, the significance of this line is not high.

        \subsection{Resolved line ratios: Gas excitation and \cishort\ as a tracer of molecular gas}
            \label{sec : resolved line ratios}
    
            \begin{figure}
                \includegraphics[scale=0.8]{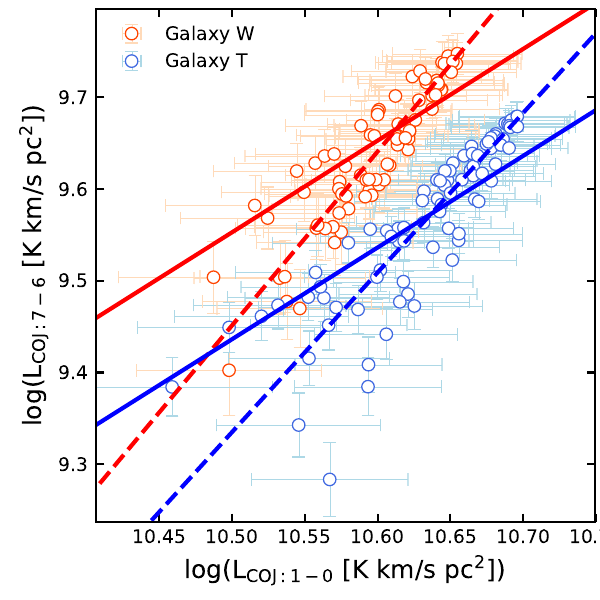} 
                \caption{
                Same as Fig.~\ref{fig : rel. of continuum luminosities}, but here we compare the emission line luminosities ($L'_{\mathrm{line}}$) of CO J:1--0 and CO J:7--6 for galaxies W and T, using kpc-scale apertures matched to the beam size of the line. The solid lines in the corresponding colour show the prediction if the galaxy integrated line flux ratio was valid over all individual apertures, and the dashed lines show the best linear fit to the datapoints.
                }
                \label{fig : co76-co10 luminosities}
            \end{figure}
    
            \begin{figure*}
                \includegraphics[scale=0.55]{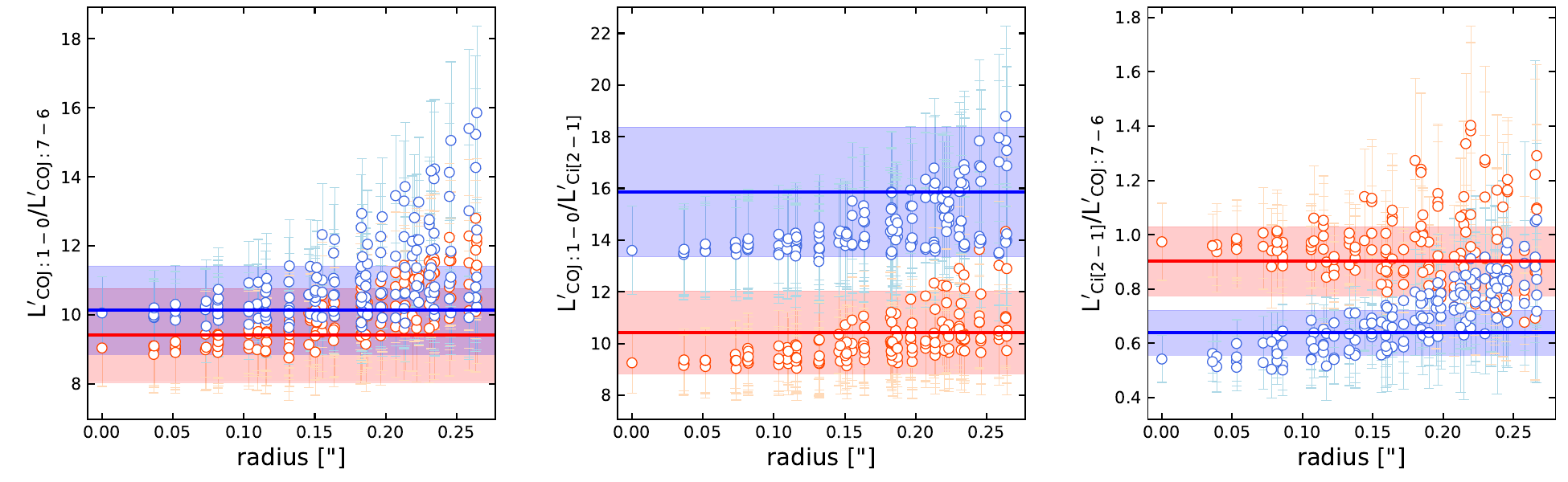}
                \caption{
                Emission line luminosity ($L'_{\mathrm{line}}$) ratios of CO J:1--0, CO J:7--6, and [C\,{\sc i}] 2--1 as a function of radius in galaxies W (red) and T (blue), measured in kpc-scale apertures matched to the beam size. In each panel the solid lines in the corresponding colour show the prediction if the galaxy integrated line flux ratio was valid over all individual apertures.
                }
                \label{fig : emission line luminosities}
            \end{figure*}
    
            Resolved line luminosities for CO J:1--0, CO J:3--2, CO J:7--6, [C\,{\sc i}] 2--1, and H$_2$O were calculated as in Sect.~\ref{sec: Galaxy-integrated spectral properties}, but instead of galaxy- and velocity-integrated line fluxes, we use velocity-integrated fluxes extracted in apertures equivalent to the synthesised beam size (FWHM), and spaced by half a synthesised beam. As before, the higher-resolution map was convolved with a Gaussian to match the resolution. Note that the CO J:3--2 images have lower resolution than the other lines. Figure~\ref{fig : co76-co10 luminosities} shows the kpc-scale resolved correlation between the CO J:1--0 and CO J:7--6 luminosities in galaxies W and T, while Fig.~\ref{fig : emission line luminosities} presents the radial trends of luminosity ratios for CO J:1--0, CO J:7--6, and [C\,{\sc i}]\,2--1. The best fits to the data from individual apertures (dashed lines) and the prediction of the galaxy-integrated ratios (solid lines) are shown for easy comparison. The galaxy-resolved CO J:7--6 and CO J:1--0 fluxes follow a relationship indistinguishable from linear, though the CO J:1--0 data are noisier, making it difficult to constrain the ratio at the lowest luminosities. The galaxy-integrated \cishort-to-CO ratios are driven by the resolved ratios in a few dominant apertures: the resolved ratios of \cishort\ to CO J:3--2 show a dependence that is significantly sub-linear. The resolved ratios of \cishort\ to CO J:7--6 are also sub-linear, though not as much as in the case of CO J:3--2: the resolved \cishort\ luminosity is typically $\sim$0.1 dex (W) or 0.2 dex (T) lower than the value that would be estimated by the galaxy-integrated line ratio. This is loosely consistent with our finding that the \cishort\ line emission is slightly more extended (at relatively low fluxes) than CO J:7--6. Further, recall that the [C\,{\sc i}] 1--0 line is typically used to trace molecular (or CO) gas. Using the \cishort\ line requires assuming an excitation temperature (where a higher temperature implies more \cishort\ emission per unit C\,{\sc i} mass). Thus, converting a CO J:7--6-to-\cishort ratio to a CO-to-C\,{\sc i} gas-mass ratio requires information on temperature and density, and the sub-linear relationship we see may therefore be mainly an effect of the physical conditions in the gas.
            
            The resolved CO J:7--6-to-CO J:3--2 ladder also shows deviations from the galaxy-integrated value, but the relatively large error bars and the highly mismatched synthesised beams of the two maps make interpretation of this figure difficult. The relationship between CO J:7--6 and \water\ shows a lot of scatter, as expected, though surprisingly the best fit to the scattered points gives an almost linear relationship between the two luminosities.
            
        \subsection{A resolved (2\,kpc-scale) Schmidt-Kennicutt relationship}
            \label{sec : S-K relationship}
    
            \begin{figure*}
                \includegraphics[scale=0.55]{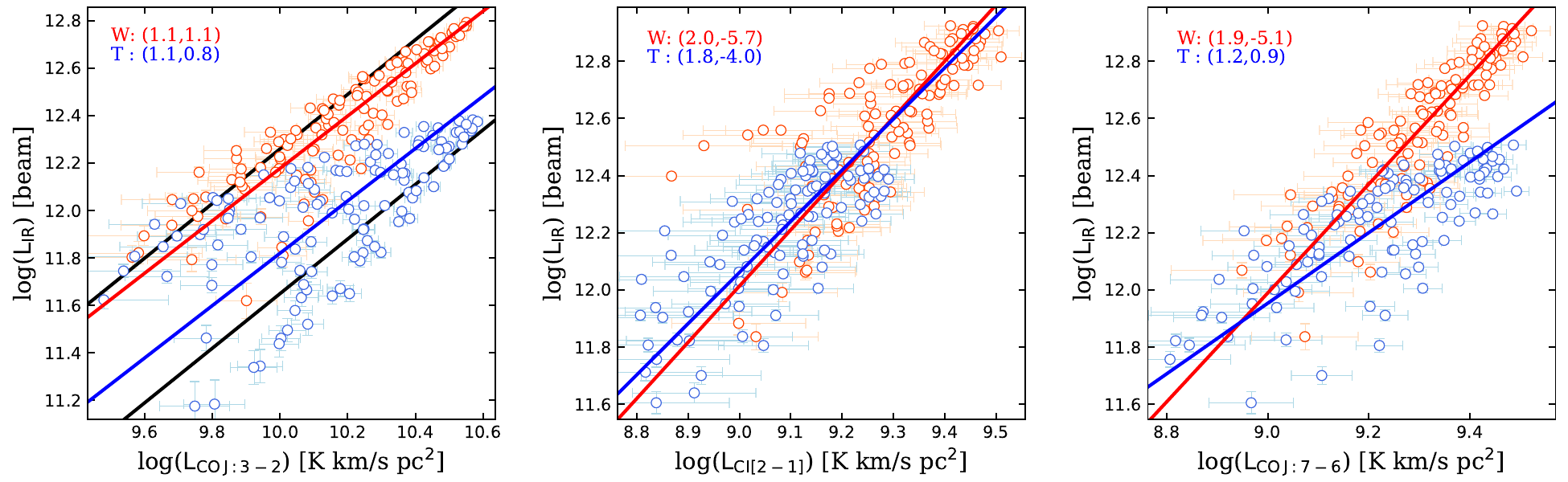}
                \caption{
                Same as Fig.~\ref{fig : rel. of continuum luminosities} but here we show the correlation between the resolved IR luminosity (estimated from the resolved rest-frame 1160 GHz continuum map; see text) and the luminosities ($L'_{\mathrm{line}}$) of the CO J:3--2, [C\,{\sc i}], and CO J:7--6 lines. In the left panel the black  lines show the galaxy-integrated relationships derived by \citet{Daddi2010} and \citet{Genzel2010} for normal star-forming (solid lower line) and `starbursting' (solid upper line) galaxies; here we assumed thermally excited gas, i.e. $L'_{\mathrm{line}_{\rm CO J:3--2}}$/$L'_{\mathrm{line}_{\rm CO J:1-0}}$ = 1.
                }
                \label{fig : IR & line luminosities}
            \end{figure*}
    
            \begin{figure*}
                \centering
                \hspace*{-0.2cm}
                \includegraphics[scale=0.85]{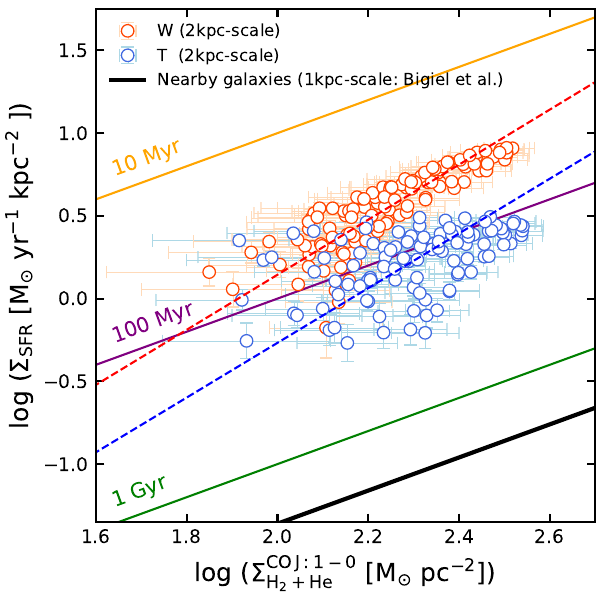}
                \includegraphics[scale=0.85]{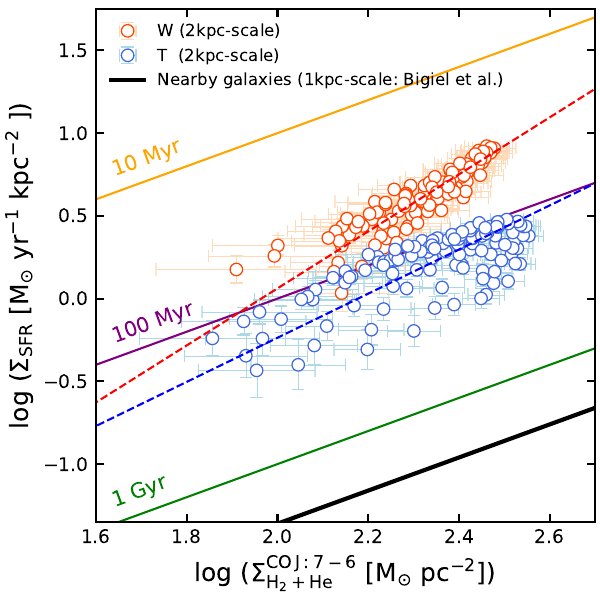} 
                \caption{
                {Resolved SK relationship, i.e.\ the relationship between the surface densities of star formation and molecular gas, for galaxies W (red points) and T (blue points). The left (right) panel uses gas surface densities derived from the CO J:1-0 (CO J:7--6) emission line (see text). The SFR surface densities are derived from the rest-frame 1160 GHz continuum emission.} Each data point was calculated over an aperture equivalent to the (FWHM) synthesised beam (roughly 2~kpc),  with a spacing of half a synthesised beam between points, i.e.\ roughly a quarter of the points are independent measurements. The solid black line (bottom-right) is the kpc-scale resolved SK relationship in nearby 'normal' galaxies from \citet{Bigiel2011}. The orange, purple and green lines delineate gas exhaustion times of (top to bottom) 10~Myr, 100~Myr y 1~Gyr. The dashed blue and red lines show the best fits to our W and T data points. In the left  panel (CO J:1-0) the slopes are 1.71 ($\pm$ 0.26) for W (intercept $-$3.2) and 1.70 ($\pm$ 0.24) for T (intercept $-$3.6). In the right panel (CO J:7-6) these are 1.72 ($\pm$ 0.20) for W (intercept $-$3.4) and 1.66 ($\pm$ 0.18) for T (intercept $-$2.9; see text). For galaxy T we used the observed (image plane) linear sizes corresponding to the synthesised beam; if the image plane is spatially stretched due to lensing then these points will move towards the top right, parallel to the gas exhaustion time lines.
                }
                \label{fig : sk relationships}
            \end{figure*}
       
            With resolved estimators of both molecular gas mass (CO J:1--0, CO J:7--6, and \cishort) and IR luminosity (rest-frame 1160~GHz continuum), it is highly relevant to test the relationship between the two, both as global quantities and as surface densities.
            
            We used the values derived by I13 for the galaxy-integrated molecular gas mass (from CO J:1--0 imaging) and total (rest-frame 8--1000~\micron) IR luminosity (from detailed SED fitting) for both W and T. Our resolved CO and rest-frame 1160~GHz continuum images are used to distribute these global values over individual resolved apertures, i.e.\ a relative rather than absolute distribution. In other words, we assume that the rest-frame 1160~GHz (260~\micron; close to the IR peak; see I13) luminosity closely traces — in a relative sense — the distribution of the IR luminosity, and that the CO J:1--0 and J:7--6 maps closely trace — in a relative sense — the distribution of molecular gas. As with the continuum maps in Fig.~\ref{fig : rel. of continuum luminosities}, we use apertures with size equal to the ALMA synthesised beam in the CO and \cishort\ maps, spaced by half a beam width. The resulting conversion factors between total continuum luminosity at 1160~GHz (in units of $\mathrm{L_{\odot}\,Hz^{-1}}$) and total $\mathrm{L_{IR}}$ (in units of $\mathrm{L_{\odot}}$) for galaxies W and T are
            \begin{equation}
                \mathrm{L_{IR} = \frac{2.414\times 10^{-14}}{\mathrm{Hz^{-1}}} L_{1160\,\mathrm{GHz}}},
            \end{equation}
            and
            \begin{equation}
                \mathrm{L_{IR} = \frac{2.256\times 10^{-14}}{\mathrm{Hz^{-1}}} L_{1160\,\mathrm{GHz}}},
            \end{equation}
            respectively. The conversions differ at the 10\% level, as each galaxy has a different global IR (from I13) to L$_{1160\,\mathrm{GHz}}$ (this work) ratio, reflecting the different shapes of their SEDs.

            We then used this value of L$_\mathrm{IR}$ to estimate the resolved SFR using the relationship of \citet{Kennicutt1998}, which assumes a Salpeter IMF: 
            \begin{equation}
                \mathrm{SFR\ \left[M_\odot\,yr^{-1}\right] = \dfrac{L_{IR}}{5.8 \times 10^9\ L_\odot}}
            \end{equation}
            Surface densities, when required, were calculated using the linear size of the synthesised beam.
    
            For molecular gas masses, we have three alternatives to derive the gas mass in each aperture, all of which use the galaxy-integrated molecular gas mass derived by I13 (using the total CO J:1--0 flux and $\alpha_{\rm CO} = 0.8$), which is distributed into apertures using our resolved emission-line imaging: 
            \begin{enumerate}
                \item directly using the resolved CO J:1--0 map of I13;
                \item using the resolved CO J:7--6 luminosity map, together with the linear fit to the relationship between the resolved CO J:7--6 and CO J:3--2 line luminosities (see Fig.~\ref{fig : emission line luminosities}), to derive an estimated high-resolution CO J:3--2 luminosity map. This is then converted to a CO J:1--0 luminosity map assuming thermally excited gas, as evidenced in Fig.~\ref{fig : co ladder}. This two-step method should be more reliable than using a direct conversion between CO J:7--6 and CO J:1--0, since we know that the CO J:7--6-to-CO J:3--2 ratio varies across the galaxy (Fig.~\ref{fig : emission line luminosities}), while the CO J:3--2-to-CO J:1--0 ratio is more likely to be constant in SMGs;
                \item using the resolved \cishort\ luminosity map, together with the galaxy-integrated \cishort-to-CO J:1--0 luminosity ratio, to create an estimated map of the CO J:1--0 luminosity.
            \end{enumerate}
    
            All three alternatives give consistent results, so here we show the results of using methods~1 and~2 above.

            We use the galaxy-integrated CO ladder of the individual galaxies (see Fig.~\ref{fig : co ladder}) to obtain the following for galaxy W: 
            $S_{\mathrm{CO\, J:7--6}}/S_{\mathrm{CO\, J:1--0}} \sim 5.52 \pm 1.64$, 
            thus $L'_{\mathrm{CO}} = (8.87  \pm 2.57) \times L'_{\mathrm{CO\, J:7--6}}$ 
            and $M_{\mathrm{mol}} = (7.10 \pm  2.06) \times L'_{\mathrm{CO\, J:7--6}}$. 
            
            For the other galaxies, the equivalent conversions are as follows:  
            Galaxy T: $M_{\mathrm{mol}} = (8.06 \pm 1.77) \times L'_{\mathrm{CO\, J:7--6}}$,  
            Galaxy M: $M_{\mathrm{mol}} = (4.85 \pm 0.87) \times L'_{\mathrm{CO\, J:7--6}}$,  
            and Galaxy C: $M_{\mathrm{mol}} = (6.06 \pm 0.13) \times L'_{\mathrm{CO\, J:7--6}}$.

            Figure~\ref{fig : IR & line luminosities} shows the dependence of the resolved IR luminosities on the CO and \cishort\ emission-line luminosities. When using CO J:3--2 on the $x$ axis (as a proxy for gas mass), both W and T follow relatively well the galaxy-integrated relationships (slope $\sim 1.2$) found by \citet{Daddi2010} and \citet{Genzel2010} \citep[cf.][]{Ivison2011} for nearby and high-redshift star-forming galaxies. In this case, most apertures in W are consistent with lying on the `luminous starburst' sequence, while most apertures in T lie between the proposed sequences for luminous starbursts and `normal star-forming' galaxies. The dependence of the IR luminosity on warm and dense gas mass (using CO J:7--6 or \cishort\ as a proxy) is significantly steeper than for CO J:3--2: the slope of the relationship is now between $\sim 1.5$ and $\sim 2.5$, and W and T are relatively indistinguishable in these plots. This is unusually steep, given that previous works have found the dependence of IR luminosity on dense gas mass (via HCN J:1--0) to be linear when galaxy-integrated quantities are considered \citep{GaoAndSolomon2004, Shimajiri2017, Oteo2017}. However, this relationship is not necessarily universal: \citet{Liu2016} find a sub-linear relationship in resolved clumps within the Galaxy. Since we are tracing both warm and dense gas with CO J:7--6, rather than only dense gas as in the case of HCN J:1--0, the steep dependence is potentially an effect of varying temperature and density across the apertures. In higher-temperature, dense regions the dependence of CO J:7--6 luminosity on SFR is steeper (the sub-mm emission would not be expected to change by a large factor, given the relatively uniform sub-mm spectral slopes observed across the galaxy). The slopes here are not the result of systematic differences in the S/N and beam shape in the individual maps. Equivalent figures constructed using data from regions between the galaxies (i.e.\ `empty sky' in the ALMA maps) show no systematic slope or offset from the origin.
            
            Converting the relationships between IR luminosity and gas (CO or \cishort) luminosity into a resolved SK relationship is trivial in our case, since all data have the same aperture size: the conversion to gas mass (rather than luminosity) and to surface density (via division by the linear aperture size) only involves a change in the axis units. Nevertheless, for clarity, we show, in a new figure (Fig.~\ref{fig : sk relationships}), the resolved SK relationship ($\mathrm{\Sigma_{SFR} \propto \Sigma_{H_2}^N}$) for galaxies W and T in the case of using CO J:1--0 (left; a cold-gas tracer) and CO J:7--6 (right; a warm/dense-gas tracer) to predict the gas surface density. Not surprisingly, the resolved SK relationships in W and T are $\sim 1$~dex higher than those seen in nearby galaxies (solid black line). Our data cover a parameter space similar to the few previous determinations of the resolved SK relationship in high-redshift galaxies \citep{Freundlich2013, Thomson2015, Hodge2015}, though we are tracing lower gas surface densities than the latter three references, reaching down to $\sim 50$~M$_\odot$\,pc$^{-2}$ in our sample. Galaxy W shows slightly higher star-formation efficiencies than T, with most apertures within 0.5~dex of a gas exhaustion time of $\sim 100$~Myr. Not surprisingly, the highest efficiencies are found in the nuclear regions, which correspond to the data points located at the upper-right end of Fig.~\ref{fig : sk relationships}.

            SK-type power-law fits to the data for W and T in Fig.~\ref{fig : sk relationships} result in different slopes depending on the specific fitting routine used. This is due both to the large spread of data points in the bottom-left quadrant of the figure and to the fork shape seen in the red points at higher gas surface densities. In all cases, the slopes are greater than 1.5. Fits to the W (red) data points typically give slopes of $1.71 \pm 0.26$ to $1.72 \pm 0.20$, while fits to the T (blue) data points give slopes of $1.70 \pm 0.24$ to $1.66 \pm 0.18$. Thus, in both W and T, the resolved cold- and warm/dense-gas SK relationships follow a power law with slope $\sim 1.7$ (dashed red and blue lines), significantly steeper than seen previously in the SK relationships of low- or high-redshift galaxies \citep[e.g.][]{Bigiel2011, Momose2013, Freundlich2013, Thomson2015, Nagy2023}.

    \section{Discussion}
        
        Our ALMA observations of the binary HyLIRG system HATLAS\,J084933.4+021443 have allowed us to perform a spatially resolved analysis of gas content, dust properties, kinematics, and star-formation activity in four distinct galaxies at $z = 2.41$. In this section we place our findings in context with previous work and discuss their implications for the star-formation law and the use of [C\,\textsc{i}](2--1) as a tracer of molecular gas in extreme systems.
        
        \subsection{The Schmidt-Kennicutt relation in HyLIRGs}
        
            The resolved SK relation presented in Sect.~3.6 shows a clear correlation between the surface densities of star-formation rate ($\Sigma_{\mathrm{SFR}}$) and molecular gas ($\Sigma_{\mathrm{gas}}$) in galaxies W and T, consistent with previous studies of main-sequence galaxies and submillimetre galaxies (SMGs) at similar redshifts \citep[e.g.][]{Tacconi2013, Genzel2015, Freundlich2019}. However, we observe significantly steeper slopes in the $\Sigma_{\mathrm{SFR}}$--$\Sigma_{\mathrm{gas}}$ relation, particularly in the nuclear regions, thereby clearly indicating enhanced star-formation efficiencies (SFEs) at high gas densities.
            
            These elevated SFEs are consistent with the nuclear starbursts observed in local ULIRGs \citep[e.g.][]{Garcia-Burillo2012}, and may suggest that gravitational instabilities and compressive turbulence dominate the ISM in these central regions \citep[e.g.][]{FederrathAndKlessen2012}. The steep slopes also support a scenario in which the SK relation in HyLIRGs transitions from a linear or sub-linear form in the disc to a super-linear regime in the nuclei, as proposed by simulations of merger-induced starbursts \citep[e.g.][]{Renaud2014}.

        \subsection{[C\,\textsc{i}] 2--1 as a tracer of warm/dense molecular gas}

            Our analysis confirms that the [C\,\textsc{i}]\,2--1 emission is spatially extended and kinematically consistent with CO J:7--6 and the dust continuum, supporting its use as a tracer of molecular gas in warm and dense environments \citep[e.g.][]{Papadopoulos2004, Jiao2017, Valentino2020}. Using the prescription from \citet{Dunne2022}, we estimate the molecular gas mass from [C\,\textsc{i}] by assuming optically thin emission and excitation temperatures consistent with the dust temperature of each galaxy, as currently best constrained.
            
            We emphasise that the [C\,\textsc{i}]\,2--1 line is particularly useful in high-redshift systems where low-J CO transitions are often inaccessible. Nonetheless, its use remains model-dependent and sensitive to the assumed excitation conditions. The agreement between CO- and [C\,\textsc{i}]-based $M_{\mathrm{H_2}}$ in our galaxies lends further support to its validity in HyLIRG-like environments. However, we caution that improved constraints on excitation temperature and density are required for more robust estimates.
            
            \medskip
            
            We further examined the relationship between [C\,\textsc{i}] and other molecular gas tracers using kpc-scale resolved line luminosities. Figure~\ref{fig : co76-co10 luminosities} shows the spatially resolved correlation between the CO J:7--6 and CO J:1--0 luminosities, whereas Fig.~\ref{fig : emission line luminosities} displays the spatially resolved radial variations in luminosity ratios including CO J:1--0, CO J:7--6, and [C\,{\sc i}]\,2--1. The resolved [C\,\textsc{i}]-to-CO J:7--6 luminosity ratios in both galaxies W and T display a roughly linear trend, with deviations of only $\sim$0.1--0.2~dex from the ratios inferred from the galaxy-integrated measurements, thereby reinforcing their reliability.
            
            This trend is loosely consistent with our finding that the [C\,\textsc{i}] emission is slightly more extended than that of CO J:7--6, particularly at lower flux levels. While [C\,\textsc{i}]\,1--0 is more commonly used as a tracer of molecular gas mass, [C\,\textsc{i}]\,2--1 can also serve this role, provided that the necessary assumptions regarding the excitation temperature are justified. Indeed, converting a CO-to-[C\,\textsc{i}] line ratio into a gas mass ratio requires knowledge of the CO excitation ladder and the physical conditions (e.g.\ temperature, density) of the emitting gas.

            Despite these model dependences, our results suggest that [C\,\textsc{i}]\,2--1 behaves as an effective tracer of warm/dense molecular gas in extreme systems such as HyLIRGs. This supports the growing body of work advocating for [C\,\textsc{i}] as an alternative to CO in high-redshift, highly star-forming environments, particularly in cases where low-J CO transitions are inaccessible or too faint to detect at present \citep[e.g.][]{Papadopoulos2004, Bothwell2017, Glover2016, Valentino2020, Dunne2022}.
            
            Overall, our findings add to the evidence that [C\,\textsc{i}] 2--1 provides a valuable window into the molecular ISM of HyLIRGs and SMGs, allowing for the estimation of gas masses and excitation conditions in regimes where classical CO-based methods can be limited.

        \subsection{Dynamical structure and gas distribution}
        
            The kinematic analysis reveals rotation-dominated discs in galaxies W and T, with some evidence of minor asymmetries or broad components suggestive of outflows or turbulence. These results, combined with the observed gas morphology and SED modelling, support the idea that HATLAS\,J084933.4+021443 is a merging system caught in an intermediate stage of interaction, where discs remain coherent but central starbursts are already active.
            
            The [C\,\textsc{i}] and CO sizes indicate compact molecular gas reservoirs in the nuclear regions, surrounded by more extended dust emission. This spatial mismatch may reflect differential obscuration or the presence of multiple ISM phases with varying temperatures and densities.

        \subsection{Uncertainties in the resolved SK relation}

            With regard to the reliability of our finding of a steep slope in the warm/dense gas SK relation, the variations in the slopes discussed above result from measurement and linear fitting errors only, without taking systematic uncertainties into account. It is thus important to examine potential systematics in the derivation of the two quantities used in the relation: SFR and gas mass.
            
            The total IR luminosity is one of the most reliable estimators of SFR \citep[e.g.][]{KennicuttAndEvans2012}—more robust than, for example, H$\alpha$-based estimators used in studies such as \citet{Freundlich2013}. In our case, the galaxy-integrated IR luminosities of W and T were derived from the comprehensive SED modelling of \citet{Ivison2013}. Ideally, SFR is derived from a combination of IR and UV luminosities (2.2\,L$_{\rm UV}$ + L$_{\rm IR}$), where the UV luminosity can be obtained via, for example, 1.5\,$\nu$L$_\nu$ at 2800\,\AA\ \citep{Bell2005}. Using the rest-frame-UV (2600\,\AA) fluxes listed in \citet{Ivison2013} for W and T (their Table~1 and Fig.~2), we find that L$_{\rm UV}$ contributes only $\sim$1--2\% of L$_{\rm IR}$ in both galaxies. The UV component is thus negligible across all apertures and can be safely ignored.

            An additional systematic in our resolved SFRs comes from using the rest-frame 250\,\micron\ ALMA map (which lies redwards of the IR peak) to distribute the total SFR among the apertures. For dust at a single known temperature—as seems to be the case in galaxy W (see Fig.~\ref{fig : rel. of continuum luminosities})—the 250\,\micron\ flux is a good proxy for IR luminosity in both absolute and relative terms \citep[e.g.][]{Orellana2017}. For T, where multiple dust temperatures are likely present, systematics are expected. If the hotter dust component(s) originates from star formation, we may underestimate the SFRs in the nuclear regions, implying an even steeper SK slope.. Conversely, if the hot dust is due to an AGN (for which {\it XMM-Newton} imaging shows no evidence; \citealt{Ivison2019}), we may overestimate the SFR, leading to a shallower intrinsic slope. Ultimately, spatially resolved maps at wavelengths short of the IR peak are required to fully assess this uncertainty.

            The systematic uncertainties in our gas mass estimates may be more substantial. However, we benefit from having three independent molecular gas tracers (CO J:1--0, CO J:7--6, and [C\,\textsc{i}]). The total molecular gas mass was estimated by \citet{Ivison2013} using CO J:1--0 and assuming $\alpha_{\rm CO} = 0.8$, and this was then apportioned into resolved apertures using our CO J:7--6 and [C\,\textsc{i}] maps. A significant change in the resolved SK slope would require a strong variation in $\alpha_{\rm CO}$ within each galaxy or substantial deviations of the CO ladder across apertures. Neither is supported by our data.

    \section{Summary and conclusions}   

        We present new resolved imaging of the continuum and emission lines in the four known galaxies of the binary HyLIRG HATLAS\,J084933.4+021443 at $z=2.41$. The new imaging allows us to further extend the comprehensive characterisation of this system presented in I13.
        
        Our main results are:
        \begin{enumerate}
            \item All four component galaxies of \src\ (W, T, M, and C) are spatially ($\sim$0\farcs3 or 2.5~kpc) resolved in CO J:7--6, \cishort, and the sub-mm continuum. Galaxies W and T are also resolved in the \water\ line.
            \item The internal kinematics of CO and \cishort\ of all four galaxies are clearly dominated by rotation.
            \item Galaxy T is significantly more extended, in gas and continuum, along its kinematic minor axis compared with Galaxy W nearby, which is likely the result of spatial magnification due to lensing.
            \item Spatially resolved sub-mm SEDs show that galaxy W is well fitted consistently with greybody emission from dust at a single temperature over the full extent of the galaxy, but galaxy T likely requires both an additional component of significantly hotter nuclear dust and additional sources of emission in the mm.
            \item We confirm that, in a rough sense, the \cishort\ line can be used as a warm/dense molecular gas tracer in such extreme systems. However, there are several caveats: an excitation temperature assumption is required for \cishort, the resolved dependence of \cishort\ on CO J:7--6 is slightly flatter than linear, and the morphology of \cishort\ and CO J:7--6 are different in W and T, with the \cishort\ slightly more extended than the CO J:7--6 emission. 
            \item We obtain an exquisite and unprecedented 2.5~kpc-scale resolved SK relationship for galaxies W and T, constructed using circular apertures of that physical size. This resolved SK relationship has been constructed with gas surface densities derived from CO J:1--0 (cold gas), CO J:7--6 (dense/warm gas), and \cishort, with all three giving consistent results. Gas exhaustion timescales are within 0.5~dex of 50~Myr for all apertures in W; those in T are about 0.4~dex slower. Both W and T follow a resolved warm/dense gas SK relationship with a power law index of $n \sim 1.7$, which is significantly steeper than the $n \sim 1$ found previously in nearby normal star-forming galaxies.
        \end{enumerate}

    \begin{acknowledgements}
        JSG acknowledges support from CONICYT project Basal AFB-170002, 
        funding  from  the CONICYT PFCHA/DOCTORADO BECAS CHILE/2019 21191147, 
        and the Predoctoral contract `Formación de Personal Investigador' from the Universidad Aut\'onoma de Madrid (FPI-UAM, 2021). 
        NN acknowledges funding support from Nucleo Milenio TITANs (NCN19$-$058).
        Funded by the Deutsche Forschungsgemeinschaft (DFG, German Research Foundation) under Germany's Excellence Strategy -- EXC-2094 -- 390783311.
        This paper makes use of the following ALMA data: ADS/JAO.ALMA\#2012.1.00334.S.
        ALMA is a partnership of ESO (representing its member states), NSF (USA) and NINS (Japan), together with NRC (Canada), MOST and ASIAA (Taiwan), and KASI (Republic of Korea), in cooperation with the Republic of Chile.
        The Joint ALMA Observatory is operated by ESO, AUI/NRAO and NAOJ.
        We thank Vicente Acu\~{n}a, Amadora Balladares and Magdalena Vilaxa for their careful read-through of the manuscript. 
    \end{acknowledgements}

    \section*{Data Availability}
        The data specifically shown in this article will be shared on reasonable request to the corresponding author.
    
    \bibliography{references}

\end{document}